\documentclass[aps, showpacs, onecolumn, preprintnumbers, 
nofootinbib, nobibnotes, superscriptaddress, longbibliography, 
amsmath, amssymb, prd, 10pt, floatfix]{revtex4-2}

\usepackage[spanish]{babel} 
\usepackage[skins,theorems]{tcolorbox}
\usepackage{color}
\usepackage{soul}
\usepackage{centernot}
\usepackage{amsfonts} 
\usepackage{graphicx} 
\graphicspath{{./Figures/}} 
\usepackage{dcolumn} 
\usepackage{siunitx}
\usepackage{bm} 
\usepackage{bbold}
\usepackage{braket}
\usepackage{esdiff} 
\usepackage[mathlines]{lineno} 
\usepackage{subfigure}
\usepackage{parskip} 
\usepackage{dsfont}
\usepackage{float} 
\usepackage{url} 
\usepackage{xcolor}
\usepackage[
    colorlinks=true,
    linkcolor=blue,
    citecolor=blue,
    urlcolor=blue,
    pdftitle={Título del Documento},
    pdfauthor={Tu Nombre},
    unicode=true
]{hyperref} 

\hypersetup{
    colorlinks=true,       
    citecolor=red,         
    linkcolor=green,       
    urlcolor=blue          
}

\showboxbreadth\maxdimen\showboxdepth\maxdimen

\newcommand{\erf}{\mathrm{erf}}
\newcommand{\erfc}{\mathrm{erfc}}

\begin{document}
\title{Thermostatistical Evaluation of Economic Activity}

\author{W. A. Rojas C.}
\email{warojasc@unal.edu.co}
\affiliation{Universidad Distrital Francisco José de Caldas, Bogot\'a, Colombia}

\author{A. Zamora V.}
\email{azamorav@udistrital.edu.co}
\affiliation{Universidad Distrital Francisco José de Caldas, Bogot\'a, Colombia}

\begin{abstract} 
We presented an analysis of Bogot\'{a}'s sports sector through the lens of  thermostatistical models  applied to economic systems. The study explores the  cross-price elasticity of Income  ($\lambda$) to assess whether sports services in Bogot\'{a} are perceived as normal or inferior goods. By examining data from the  Sports Satellite Account of Bogot\'{a} (CSDB)  from 2018 to 2022, we  found  that the demand for sports services was highly elastic, especially in years when the economy improved, implying that these services are viewed as normal or luxury goods.

Additionally, the  partition function,  entropy, and  heat capacity  of the economic system are calculated, revealing that the entropy is consistent with the  Boltzmann Principle. This indicates a strong correlation between the system's microstates and its macroeconomic state, supporting the statistical thermodynamic framework used.

The study also applies  geometrothermodynamics  to evaluate the system's stability, utilizing  Kretschmann  and  Ricci scalars  to uncover economic singularities or crises. These singularities coincide with the pandemic period, reflecting its disruptive impact on the sector. This integration of geometric thermodynamics offers a deep interpretation of the system’s stability, revealing how external shocks, such as COVID-19, cause measurable disturbances in the economic structure.

The analysis shows that Bogot\'{a}'s sports sector responds elastically to changes in GDP, and its stability is influenced by various macroeconomic factors. However, the decrease in the heat capacity of the system as the economic temperature rises suggests potential growth limitations, making further research essential to fully understand the sector's long-term prospects.
\end{abstract} 
\maketitle

\section{Introduction}\label{sec1}

Econophysics is a discipline derived from the convergence between statistical physics and economics, employing advanced physical methods to analyze and model the inherent complexity of economic systems. This approach facilitates the interpretation of financial and economic phenomena as complex systems, leveraging the analysis of large datasets. The integration of methods from statistical physics can improve existing economic models and offer more accurate predictions of economic phenomena. The growing availability of economic data presents opportunities to discover new empirical laws and validate existing ones \cite{stanley1999}.

The integration of thermodynamics with economic theory is crucial for developing a coherent economic theory that considers sustainability. Economic processes follow the laws of thermodynamics, especially the conservation of mass and energy and the law of entropy. In this context, the concepts of strong and weak sustainability are analyzed. Strong sustainability requires that natural and economic capital do not decrease, while weak sustainability assumes perfect substitutability between these capitals, which is considered unrealistic. The second law of thermodynamics limits the possibility of a completely circular and sustainable economy due to the entropic nature of material use. The concept of exergy, as a measure of the value of energy in terms of work, can be used to assess the environmental impact and sustainability of different processes and areas \cite{Ciegis2008}.

Santos et al. \cite{Santos2016} present various kinetic models for the distribution of money and wealth that apply concepts from statistical mechanics and quantum physics. The study examines how both constant and differentiated saving propensity affect the distributions of money and wealth, using the Gini index to quantify the degree of inequality. An analogy is drawn between economic systems and ideal gases, where economic transactions are modeled as collisions between particles, conserving an economic quantity analogous to kinetic energy.

Rawlins et al. \cite{rawlings2004} develop an economic statistical thermodynamics framework where entropy naturally emerges. Using this framework, they derive Pareto's law for income distribution, based on the conservation of the sum of the logarithms of income in the high-income group. Thus, they establish a relationship between the Pareto exponent and what they call economic temperature, empirically demonstrating that different U.S. states exhibit the same economic temperature, suggesting the existence of a quasi-equilibrium in these economic systems.

Quevedo et al. \cite{Quevedo2011} estimate an econophysics model that takes into account the microeconomic and macroeconomic parameters of a system. In this formulation, the free money function and Pareto's law can be derived. This approach allows for the estimation of entropy $S$  and heat capacity $C$.
\begin{equation}
C=T\frac{\partial S}{\partial T},
\label{aqn1400}
\end{equation}
Thus, the divergences in \eqref{aqn1400} are associated with phase transitions and crises in economic systems \cite{Quevedo2011, Quevedo2023}.

Quevedo shows the application of  geometrothermodynamics (GTD)  in econophysics to describe the behavior of economic systems. GTD, which uses concepts from differential geometry, is applied to economic systems by modeling them as thermodynamic systems. The research focuses on the  Boltzmann-Gibbs  and  Pareto distributions , which represent different population sectors $95\%$ with medium and low wealth, and $5\%$ with high wealth. The study demonstrates that the Boltzmann-Gibbs sector does not exhibit phase transitions, while the Pareto sector, characterized by strong economic interactions, presents significant  curvature singularities , interpreted as  economic crises \cite{Quevedo2023}. The article also derives the  partition function ,  entropy , and  heat capacity  of the system, showing that the entropy complies with the  Boltzmann Principle . This provides a robust statistical and thermodynamic foundation for the model. By associating Riemannian metrics to these distributions, the authors reveal that the curvature in the equilibrium space helps identify financial crises as  phase transitions \cite{Quevedo2023}.

This research is organized as follows: Section \ref{sec2} presents a review of elements of microeconomics and the Satellite Account of Sports in Bogot\'{a} \cite{dane2023}. In Section \ref{sec3}, a review of the statistical-physical model applied to economics is provided \cite{Quevedo2011, Quevedo2016}. Subsequently, a brief review of elements of Geometrothermodynamics (GTD) for economic systems is presented in Section \ref{sec4}. Section \ref{sec5} offers a thermal description of the Satellite Account of Sports (CSDB). This section is dedicated to applying GTD to the economic sector of interest in Section \ref{sec6}. Finally, discussions and conclusions are presented in Section \ref{sec7}.

\section{The Satellite Account of Sports in Bogot\'{a} }\label{sec2}
\subsection{Elements of Microeconomics}
Microeconomics focuses on the individual behavior of economic agents, such as consumers, firms, and workers, and how their decisions affect the supply and demand for goods and services. When considering microeconomics, supply and demand must be taken into account. Thus, demand is the quantity of goods or services that consumers are willing to purchase at different prices. The Law of Demand states that when the price decreases, the quantity demanded increases.

Similarly, supply corresponds to the quantity of goods or services that producers are willing to sell at different prices. The Law of Supply indicates that when the price increases, the quantity supplied also increases \cite{arya1997}.

Demand elasticity measures the sensitivity of the quantity demanded to changes in price. Elastic demand responds significantly to price changes, while inelastic demand changes little. Supply elasticity measures the sensitivity of the quantity supplied to changes in price \cite{arya1997}.

\subsection{The Satellite Account of Sports in Bogot\'{a} (CSDB)}
The Satellite Account of Sports in Bogot\'{a} (CSDB)\footnote{For its Spanish acronym.}\cite{dane2023} is a fundamental tool for measuring and analyzing the economic impact of sports-related activities in the Colombian capital. This account, developed by DANE \cite{dane2023} in collaboration with the Mayor's Office of Bogot\'{a} \cite{bogota2023} and the District Institute of Recreation and Sports (IDRD) \cite{idrd2023}, provides data on the value added generated by the sports sector and its contribution to the city's GDP. According to the results presented, sports economic activities represented, on average, $1.11\%$ of the total value added in Bogot\'{a} between 2018 and 2022, demonstrating their relevance in the local economy. The CSDB encompasses various activities, ranging from manufacturing and commerce to services related to sports, offering a comprehensive view of the sector.

This detailed analysis allows policymakers and private-sector stakeholders to make informed decisions to promote the development and growth of the sports industry in Bogot\'{a}. Additionally, the CSDB facilitates the comparison of the economic performance of the sports sector over time and its evolution against other sectors of the economy, which is crucial for evaluating the impact of policies and strategies aimed at promoting sports and physical activity in the city \cite{dane2023}.

\begin{figure}[htbp]
\centering
		\includegraphics[width=0.7\textwidth]{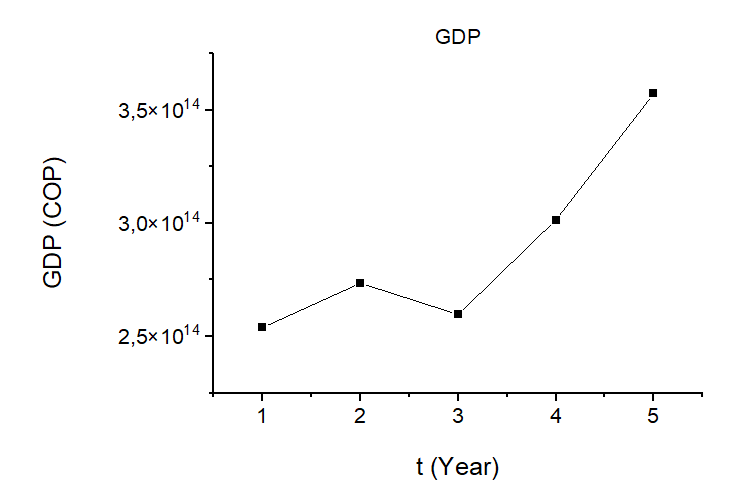}
\caption{GDP for Bogot\'{a} City during the period 2018-2022.}
\label{WA01}
\end{figure}
In Figure \ref{WA01}, the behavior of the Gross Domestic Product (GDP) for the city of Bogot\'{a} during the period 2018-2022 is presented according to the DANE report \cite{dane2023}. A least-squares adjustment for GDP allows us to obtain
\begin{equation}
f(\lambda_{2})= \alpha_{2}\lambda_{2}^{3}+\beta_{2}\lambda_{2}^{2}+\gamma_{2}\lambda_{2}+\delta_{2},
\label{WA11}
\end{equation}
where $\alpha_{2}=4*10^{12}$, $\beta_{2}=-2*10^{16}$, $\gamma_{2}=5*10^{19}$ and $\delta_{2}=3*10^{22}$ with a correlation coefficient of $R^{2}=0.96$.

\begin{figure}[htbp]
\centering
		\includegraphics[width=0.7\textwidth]{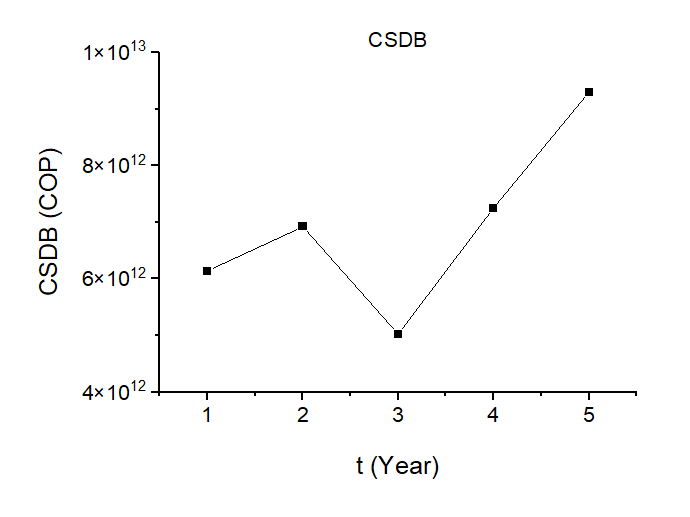}
\caption{CSDB  for Bogot\'{a} City during the period 2018-2022.}
\label{WA10}
\end{figure}
In Figure \ref{WA10}, the behavior of the CSDB for the city of Bogot\'{a} during the period 2018-2022 is presented according to the DANE report \cite{dane2023}. It is possible to calculate the cross elasticity $\lambda$ with the available data. Therefore, $\lambda$ measures the sensitivity of one variable (CSDB) with respect to changes in another variable (GDP), as \cite{dowling1997, arya1997, hoy2001, hoffmann2007}.
\begin{equation}
\lambda=\frac{\frac{\Delta CSDB}{\left\langle CSDB\right\rangle}}{\frac{\Delta GDP}{\left\langle GDP\right\rangle}},
\label{aqn1500}
\end{equation}
where
\begin{equation}
\Delta CSDB=CSDB_{i+1}-CSDB_{i},
\label{aqn1510}
\end{equation}

\begin{equation}
\Delta GDP=GDP_{i+1}-GDP_{i},
\label{aqn1520}
\end{equation}

\begin{equation}
\left\langle CSDB\right\rangle=\frac{CSDB_{i+1}+CSDB_{i}}{2}
\label{aqn1530}
\end{equation}

\begin{equation}
\left\langle GDP\right\rangle=\frac{GDP_{i+1}+GDP_{i}}{2}
\label{aqn1540}
\end{equation}

\begin{figure}[htbp]
\centering
		\includegraphics[width=0.7\textwidth]{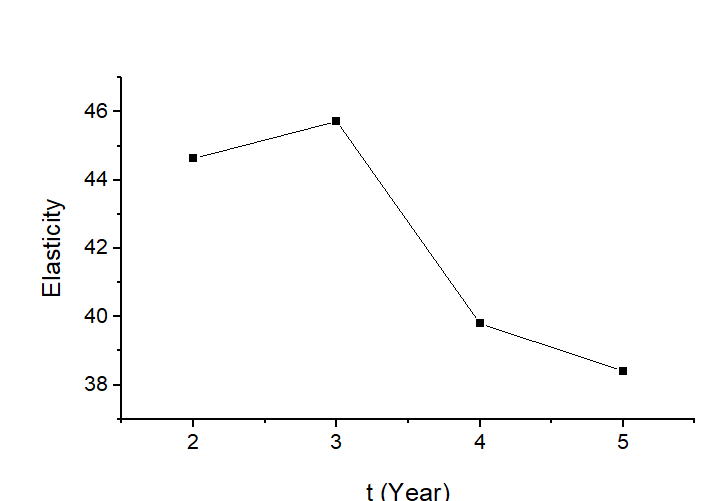}
\caption{Behavior of $\lambda$ for Bogot\'{a} City during the period 2018-2022.}
\label{WA20}
\end{figure}
Figure \ref{WA20} shows the behavior of the elasticity coefficient $\lambda$ for Bogot\'{a} City during the period 2018-2022.  A least-squares adjustment for the data in Figure \ref{WA20} allows us to obtain the polynomial $m(\lambda)$
\begin{equation}
m(\lambda)=\alpha \lambda^{3}+\beta \lambda^{2}+\gamma\lambda+\delta,
\label{aqn1550}
\end{equation}

where $\alpha=1.9$, $\beta=-1.1*10^{4}$, $\gamma=2.0*10^{6}$, $\delta=2.0*10^{10}$ with a correlation coefficient of $R^{2}=1$. 

Thus, $\lambda$ defined in \eqref{aqn1500} is a cross-income elasticity that measures the sensitivity of the quantity demanded or supplied of a good or service in response to changes in income or another economic variable, such as GDP.

In this specific case, the elasticity of the supply of sports spending in Bogot\'{a} is being measured in response to changes in Bogot\'{a}'s GDP. Characteristics of $\lambda$: if the increase in GDP results in an increase in sports spending, the elasticity is positive, suggesting that sports is a normal good.

Similarly, if an increase in GDP results in a decrease in sports spending, the elasticity is negative, suggesting that sports may behave as an inferior good.  Also, $\lambda > 1$ indicates that the quantity supplied is highly sensitive to changes in GDP. In contrast, $\lambda < 1$ is inelastic, indicating that the quantity supplied is less sensitive to changes in GDP.

In the context of Supply, it would be observed how changes in GDP affect the ability or willingness to offer sports-related services in Bogot\'{a}. Similarly, for Demand, it would be observed how changes in GDP affect the consumption of sports services in Bogot\'{a}.

Income elasticity provides information on how the supply or demand for sports services changes in response to changes in GDP, which can indirectly affect the average spending per market agent \cite{dane2023}.

\section{Elements of statistical thermodynamics applied to economic systems.}\label{sec3}
Quevedo \textsl{et al.} \cite{Quevedo2011, Quevedo2016, Quevedo2023} present a physical-statistical model applied to economics, where a quantity $M$ is conserved over a period of time $t$
\begin{equation}
\frac{dM}{dt}=0.
\label{eqn00}
\end{equation}
Let $N$ be the number of market agents competing for a share of the money \cite{Santos_2016}
\begin{align}
    	M&=m_{1}+m_{2}+\ldots+m_{i}   
\notag\\
	&=\sum^{N}_{i}m_{i}.
&\hspace{0.3cm}
\label{eqn10}
\end{align}
In the case of the canonical ensemble, the distribution density of money is
\begin{equation}
\rho(m)\propto e^{-m/T},
\label{eqn20}
\end{equation}
where $T=\left\langle E\right\rangle$, which corresponds to the average energy per market agent (money), $m=m(\lambda)$
 corresponds to the money function and is the amount of money that a market agent can obtain depending on $\lambda_{1},\lambda_{2},\ldots$
 microeconomic parameters. Therefore, \eqref{eqn20} reduces to
\begin{equation}
\rho(\bar{\lambda})=\frac{e^{-m(\bar{\lambda})/T}}{Z(T,\bar{x})},
\label{eqn30}
\end{equation}
where $Z(T,\bar{x})$ it is the partition function \cite{Quevedo2011,Quevedo2016,Quevedo2023,Santos_2016,RAWLINGS2004643,alonso1976fisica,reif1996fisica,landau1988fisica,greiner2012thermodynamics}
\begin{equation}
Z(T,\bar{x})=\int e^{-m(\bar{\lambda})/T}d\bar{\lambda},
\label{eqn40}
\end{equation}
where $\bar{\lambda}$ is the entire set of microparameters, $\bar{x}$ is the entire set of macroeconomic parameters. Thus, the average value of any parameter $g=g(\bar{\lambda})$ is 
 \begin{align}
    	\left\langle g\right\rangle&= \int g \rho d\bar{\lambda}  
\notag\\
	&=\frac{1}{Z(T,\bar{x})}\int g e^{-m(\bar{\lambda})/T}d\bar{\lambda}.
\label{eqn50}
\end{align}
From the above, the average amount of money per agent
\begin{equation}
\left\langle m\right\rangle= \int m \rho d\bar{\lambda},
\label{eqn60}
\end{equation}
Consequently
\begin{equation}
d\left\langle m \right\rangle=\int m d\rho d\bar{\lambda}+\int \rho dm d\bar{\lambda}.
\label{eqn70}
\end{equation}
Thus, in \eqref{eqn70}, it is defined
\begin{equation}
\left\langle dm\right\rangle=\int \rho dm d\bar{\lambda},
\label{eqn80}
\end{equation}
then \eqref{eqn70} reduces to
\begin{equation}
d\left\langle m \right\rangle=\int m d\rho d\bar{\lambda}+\left\langle dm\right\rangle.
\label{eqn90}
\end{equation}
Taking \eqref{eqn30}, it is rewritten
\begin{equation}
m(\bar{\lambda})=-T\left[\ln\left|\rho(\bar{\lambda})\right|+\ln\left|Z(T,\bar{x})\right|\right].
\label{eqn100}
\end{equation}
hus, for \eqref{eqn100}, it holds that
\begin{equation}
\int \rho(\bar{\lambda})d\bar{\lambda}=1,\,\,\,d\int \rho(\bar{\lambda})d\bar{\lambda}=0.
\label{eqn110}
\end{equation}
Taking into account the term 
\begin{align}
    \int m d\rho d\bar{\lambda}	&=  -\int T\left[\ln\left|\rho(\bar{\lambda})\right|+\ln\left|Z(T,\bar{x})\right|\right] d\rho d\bar{\lambda} 
\notag\\
	&=-\int T\left[\ln\left|\rho(\bar{\lambda})\right|\right]d\rho d\bar{\lambda}
\label{eqn120}
\end{align}
where the second term is null due to \eqref{eqn110}.

Let the entropy $S$ of a thermodynamic system be defined as

\begin{align}
    	S&=  \left\langle -\ln\left|\rho(\bar{\lambda})\right|\right\rangle 
\notag\\
	&=-\int \ln\left|\rho(\bar{\lambda})\right|\rho(\bar{\lambda}) d\bar{\lambda},
\label{eqn130}
\end{align}
therefore, its differential $dS$ is  

\begin{equation}
ds=-\int \ln\left|\rho(\bar{\lambda})\right|d\rho(\bar{\lambda})d\bar{\lambda}.
\label{eqn140}
\end{equation}

Let us consider again equation \eqref{eqn80}, which can be rewritten as

\begin{align}
    	\left\langle -\frac{\partial  m}{\partial x_{i}}\right\rangle&= \int\left[-\frac{\partial m}{\partial x_{i}}\right]\rho{\bar{\lambda}}d  \bar{\lambda}
\notag\\
	y_{i}&=\int\left[-\frac{\partial m}{\partial x_{i}}\right]\rho{\bar{\lambda}}d  \bar{\lambda},
\label{eqn150}
\end{align}
Thus, for \eqref{eqn150}, it is given that  $\left[-\frac{\partial m}{\partial x_{i}}\right]$  corresponds to the variation in the amount of money $m$  per market agent with respect to market factors  $x_{i}$. Then
\begin{equation}
-y_{i}dx_{i}=\int \rho(\bar{\lambda})dm d\bar{\lambda}.
\label{eqn160}
\end{equation}
Consequently
\begin{align}
    	\left\langle dm\right\rangle&= -\sum_{i}^{N}y_{i}dx_{i}.  
\notag\\
	&=\int \rho(\bar{\lambda})dm d\bar{\lambda}.
\label{eqn170}
\end{align}
Taking \eqref{eqn90} according to \eqref{eqn140} and \eqref{eqn170}, the First Law of Thermodynamics for an economic system is obtained \cite{Quevedo2011, RAWLINGS2004643}
\begin{equation}
d\left\langle m\right\rangle=TdS-\sum_{i}^{N}y_{i}dx_{i},
\label{eqn180}
\end{equation}
where the term of heat is interpreted $dQ=TdS$ and the term of work  $dW=\sum_{i}^{N}y_{i}dx_{i}$.

It is useful to explicitly calculate the entropy. $S$,  from \eqref{eqn130}. Taking \eqref{eqn100}
\begin{equation}
\ln\left|\rho(\bar{\lambda})\right|=\frac{m(\bar{\lambda})}{T}+\ln\left|Z(T,\bar{x})\right|.
\label{eqn190}
\end{equation}
Then
\begin{equation}
S=\int\left[\frac{m(\bar{\lambda})}{T}+\ln\left|Z(T,\bar{x})\right|\right]\rho(\bar{\lambda})d\bar{\lambda}.
\label{eqn200}
\end{equation}
The free money function of an economic system can be defined analogously to Helmholtz free energy \cite{landau1988fisica,greiner2012thermodynamics}
\begin{equation}
f=\left\langle m\right\rangle-TS
\label{eqn210}
\end{equation}
\begin{equation}
f=-T\ln\left|Z(T,\bar{x})\right|.
\label{eqn220}
\end{equation}
From \eqref{eqn210}, it is defined
\begin{equation}
S=-\left(\frac{\partial f}{\partial T}\right)_{\left\langle m\right\rangle},
\label{eqn230}
\end{equation}
\begin{equation}
y_{i}=-\left(\frac{\partial f}{\partial x_{i}}\right)_{TS}.
\label{eqn240}
\end{equation}
Once the entropy $S$ is known, the heat capacity can be determined as 
\begin{equation}
C=T\frac{\partial S}{\partial T}.
\label{eqn241}
\end{equation}
\section{Elements of geometric thermodynamics for economic systems}\label{sec4}
Geometric thermodynamics  (GTD) consists of introducing a metric in the equilibrium space $\mathcal{E}$, such that the points ($P\in\mathcal{E}$) represent all possible equilibrium states of the system \cite{Quevedo2023,Quevedo2007,Larranaga2011,valdes2016interpretacion,pineda2019geometrotermodinamica}.

Define $\Phi(E^{a})$ with $a=1,\ldots n$ as the fundamental equation of the system, where $\Phi$ is the thermodynamic potential, $E^{a}$  are the extensive variables that correspond to the coordinates of $\mathcal{E}$ and $n$ is the number of degrees of freedom. Thus, the Hessian metric is
 \begin{equation}
g^{H}=\frac{\partial^{2}\Phi}{\partial E^{a}\partial E^{B}}dE^{a}dE^{b}.
\label{gtd00}
\end{equation}
Consequently, it is considered that the equilibrium space has a Riemannian manifold structure; therefore $\Phi=U$ is the Weinhold metric and $\Phi=-S$ is the Ruppeiner metric.

Then, the thermal description of a physical system must be independent of the chosen thermodynamic potential; this characteristic is called Legendre invariance. Thus, metrics $g$ are defined over $\mathcal{E}$ that comply with this invariance principle. A defect of 
\eqref{gtd00} is that it does not adhere to the principle of Legendre invariance. Nevertheless, GTD offers a set of metrics that satisfy this invariance.
\begin{equation}
g^{I}=\beta_{\Phi}\Phi\delta^{c}_{a}\frac{\partial^{2}\Phi}{\partial E^{b}\partial E^{c}},
\label{gtd20}
\end{equation}
\begin{equation}
g^{II}=\beta_{\Phi}\Phi\eta^{c}_{a}\frac{\partial^{2}\Phi}{\partial E^{b}\partial E^{c}},
\label{gtd30}
\end{equation}
\begin{equation}
g^{III}=\sum^{n}_{a=1}    \left[ \delta_{ab}E^{d}\frac{\partial \Phi}{\partial E^{a}}\right]\delta^{ab}\frac{\partial^{2}\Phi}{\partial E^{b}\partial E^{c}}dE^{a}dE^{c},
\label{gtd40}
\end{equation}
where $\delta^{c}_{a}=\mbox{diag}(1,\ldots ,1)$, $\eta^{c}_{a}=\mbox{diag}(-1,1,\ldots ,1)$ y $\beta_{\Phi}$ is the degree of homogeneity 
of $\Phi$ \cite{valdes2016interpretacion,pineda2019geometrotermodinamica}.

\section{Thermal Description of the Sports Satellite Account (CSDB)}\label{sec5}
This study aims to analyze the sports sector of Bogot\'a City \cite{dane2023}  using methods from thermodynamics applied to economics \cite{Quevedo2011, Quevedo2023}. Consequently, the partition function $Z(T, \lambda)$ is given according to \eqref{aqn1550} and \eqref{eqn40}. 
\begin{equation}
Z(T, \lambda)=\frac{\bar{\Lambda}}{\Lambda}\int^{\infty}_{0}\exp\left[-\frac{\alpha \lambda^{3}+\beta \lambda^{2}+\gamma\lambda+\delta}{T}\right]d\lambda,
\label{eqn250}
\end{equation}
where $\bar{\Lambda}=\Lambda_{1}\Lambda_{2}\,\ldots$ corresponds to the macroeconomic parameters of the system under consideration. The presence of the polynomial in the argument of the exponential function makes the integral \eqref{eqn250} have no known analytical solution. However, depending on the values of $\alpha$, $\beta$, $\gamma$, and $\delta$, specific approximations can be made.

A quadratic approximation can be considered
\begin{equation}
\alpha=0.
\label{eqn255}
\end{equation}
This allows \eqref{eqn250} to be reduced to
\begin{widetext} 
\begin{align}
    	Z(T, \lambda)&=\frac{\bar{\Lambda}}{\Lambda}\int^{\infty}_{0}\exp\left[-\frac{\beta \lambda^{2}+\gamma\lambda+\delta}{T}\right]d\lambda  
\notag\\
	&=\frac{\bar{\Lambda}}{\Lambda}\int^{\infty}_{0}\exp\left[\frac{-\beta\left(\lambda+\frac{\gamma}{2\beta}\right)^{2}-\frac{\gamma^{4}}{4\beta}+\delta}{T}\right]d\lambda
&\hspace{0.3cm}
\notag\\
	&=\frac{\sqrt{\pi}}{2\beta}\frac{\bar{\Lambda}}{\Lambda}\exp\left[-\frac{1}{T}\left(\frac{\gamma^{2}}{4\beta}+\delta\right)\right] \left[T\sqrt{\frac{\beta}{T}}-\sqrt{T\beta} \,\,*\erfc \left[\frac{\gamma^{2}}{2\sqrt{T\beta}} \right] \right]
&\hspace{0.3cm}
\label{eqn260}
\end{align}
\end{widetext}
Where $\erfc(z)$ and $\erf(z)$ are the complementary error function and the error function, respectively \cite{spiegel1999mathematical}
\begin{equation}
\erfc(z)=1-\erf(z)
\label{eqn270}
\end{equation}
\begin{equation}
\erf(z)=\frac{2}{\pi}\int^{z}_{0}e^{-t^{2}}dt.
\label{eqn280}
\end{equation}
An analytical evaluation of \eqref{eqn260} is beyond the scope of the present study; however, it is illustrative to attempt a computational alternative using Mathematica, which may help estimate the thermal behavior of CSDB.

\begin{figure}[htbp]
\centering
		\includegraphics[width=0.5\textwidth]{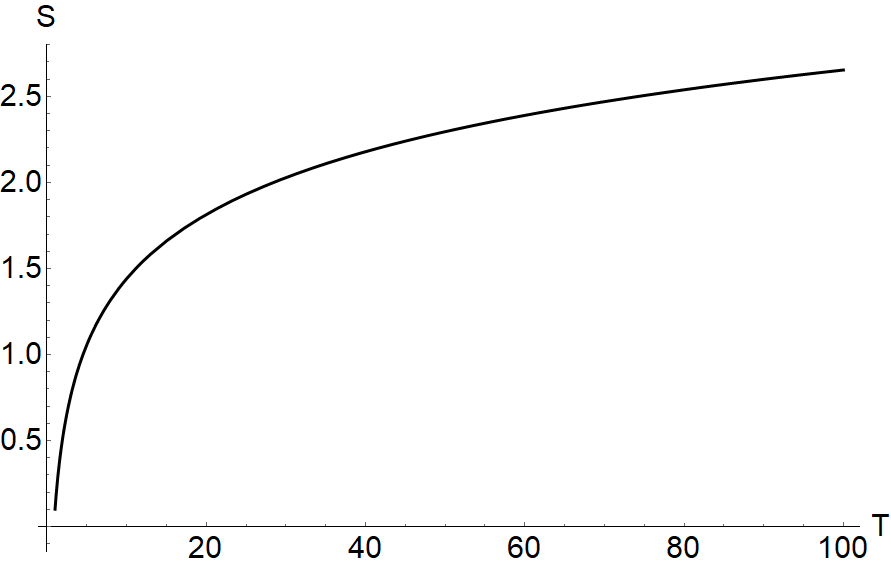}
\caption{Behavior of entropy $S$ according to the partition function $Z(T, \lambda)$, \eqref{eqn260}}
\label{WA30}
\end{figure}

In Figure \ref{WA30}, the behavior of entropy $S(T)$ for the CSDB is presented according to the partition function \eqref{eqn260}. A logarithmic behavior of entropy is observed. This entropy is thermal and aligns with Boltzmann's principle, where entropy  is interpreted as proportional to the logarithm of the number of microstates compatible with  given a macrostate, $S\propto \ln \left| \Omega\right|$. This point is significant as we consider the CSDB as a physico-statistical phenomenon. That is, the CSDB possesses thermal entropy, and according to the second law of thermodynamics, it is associated with a temperature $T$. Here, it is important to distinguish the concept of physical temperature \(T_{\text{Ph}}\) from economic temperature \(T_{\text{Eco}}\), as the former is associated with the motion of particles in a physical system and their average velocity, \(\left\langle v \right\rangle\), while the latter is interpreted as the average amount of money, \(\left\langle m \right\rangle\), held by the different agents in the CSDB \cite{rawlings2004}.

\begin{figure}[htbp]
\centering
		\includegraphics[width=0.5\textwidth]{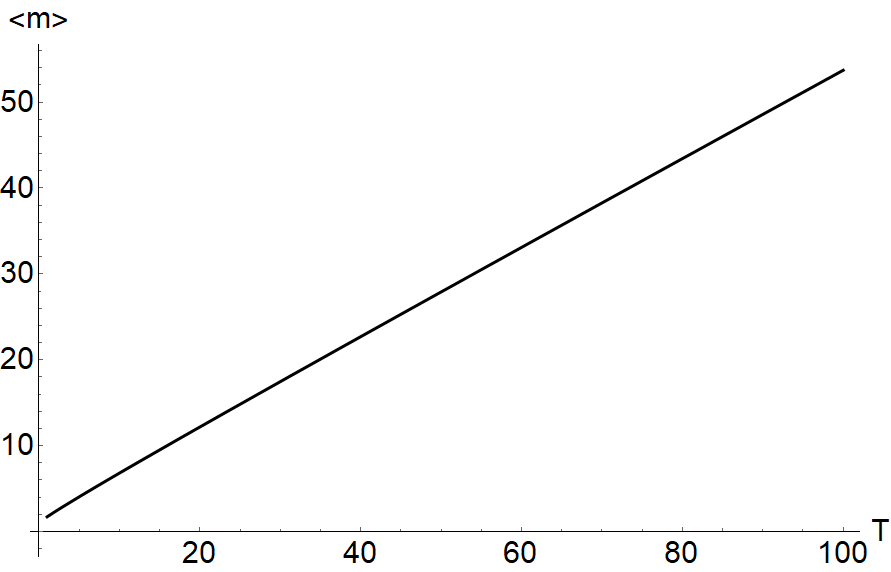}
\caption{Behavior of $\left\langle m \right\rangle$ according to the partition function $Z(T, \lambda)$, \eqref{eqn260}.}
\label{WA40}
\end{figure}
Figure \ref{WA40} shows the behavior of $\left\langle m \right\rangle(T)$, where a linear relationship between the two analyzed variables is observed.
\begin{figure}[htbp]
\centering
		\includegraphics[width=0.5\textwidth]{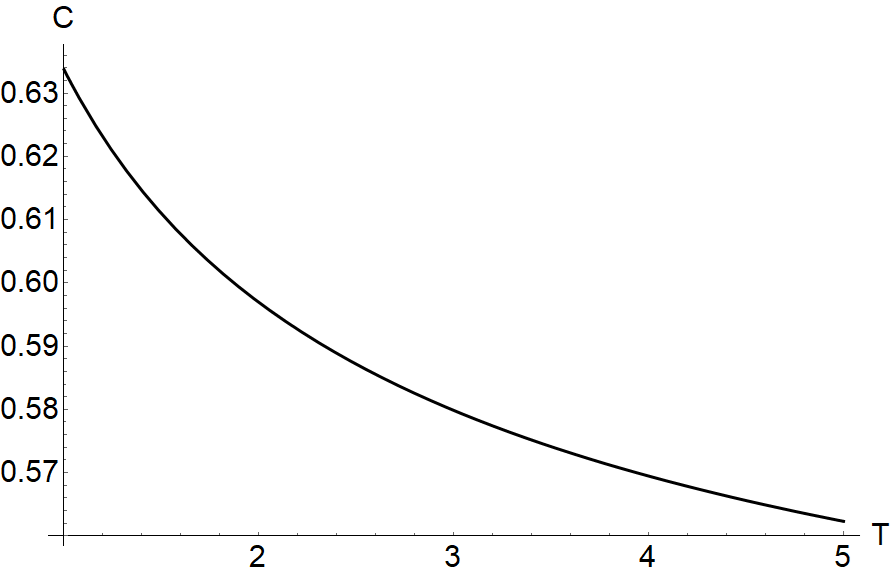}
\caption{Behavior of the heat capacity $C$ according to the partition function $Z(T, \lambda)$, \eqref{eqn260}.}
\label{WA50}
\end{figure}

Figure \ref{WA50} shows the behavior of $C(T) \propto \frac{1}{T}$.

Consider again \eqref{aqn1550}, with the cubic approximation
\begin{equation}
\alpha \lambda^{3}\gg \beta \lambda^{2}+\gamma \lambda+\delta.
\label{eqn290}
\end{equation} 
Therefore, the partition function
\begin{align}
	Z(T, \lambda)&=\frac{\bar{\Lambda}}{\Lambda}\int^{\infty}_{0}\exp\left[-\frac{\alpha \lambda^{3}}{T}\right]d\lambda  
\notag\\
	&=\frac{\bar{\Lambda}}{\Lambda}\sqrt[3]{\frac{T}{\alpha}}\Gamma\left[\frac{4}{3}\right].
&\hspace{0.3cm}
\label{eqn300}
\end{align}
It is possible to obtain an analytical solution as reported by \cite{Quevedo2011,Quevedo2023}. The free money function, based on \eqref{eqn220}
\begin{equation}
f=-T\ln\left|\frac{\bar{\Lambda}}{\Lambda}\sqrt[3]{\frac{T}{\alpha}}\Gamma\left[\frac{4}{3}\right]\right|.
\label{eqn310}
\end{equation}
Similarly, the entropy is obtained from \eqref{eqn230}
\begin{equation}
S=\frac{1}{3}+\ln\left|\frac{\bar{\Lambda}}{\Lambda}\sqrt[3]{\frac{T}{\alpha}}\Gamma\left[\frac{4}{3}\right]\right|.
\label{eqn320}
\end{equation}

\begin{figure}[htbp]
\centering
		\includegraphics[width=0.5\textwidth]{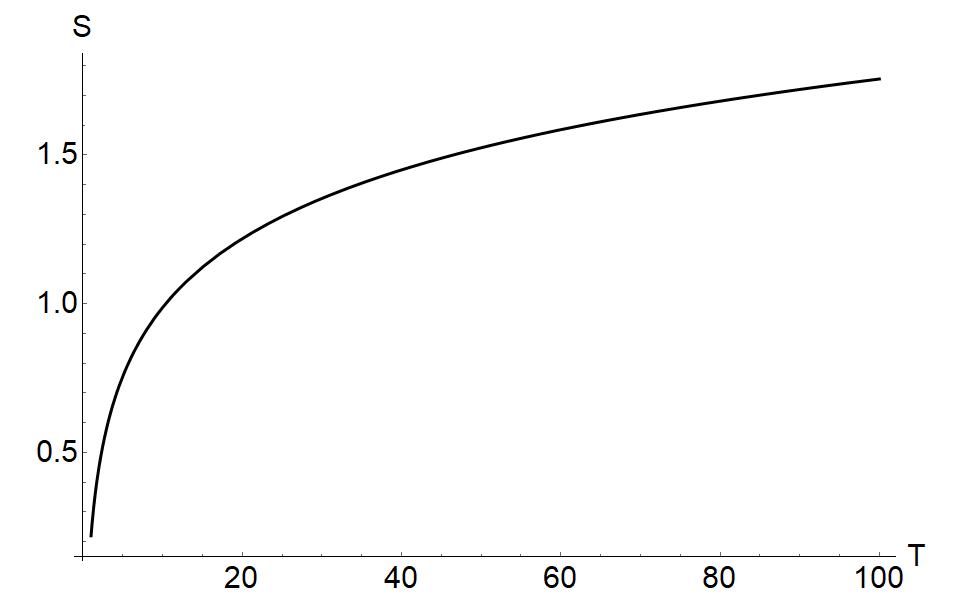}
\caption{Behavior of the entropy $S$  according to the partition function  $Z(T,\lambda)$, \eqref{eqn300}.}
\label{WA60}
\end{figure}
In Figure \ref{WA60}, the behavior of the entropy $S$ is presented according to the partition function $Z(T,\lambda)$ in \eqref{eqn300}. It is observed that the entropy exhibits a logarithmic behavior $S \propto \ln |\Omega|$, similar to Figure \eqref{WA30}.

Similarly, from \ref{eqn210}, the average value of money per agent, $\langle m \rangle$, can be obtained as:
\begin{equation}
\left\langle m\right\rangle=\frac{T}{3}.
\label{eqn321}
\end{equation}
\begin{figure}[htbp]
\centering
		\includegraphics[width=0.5\textwidth]{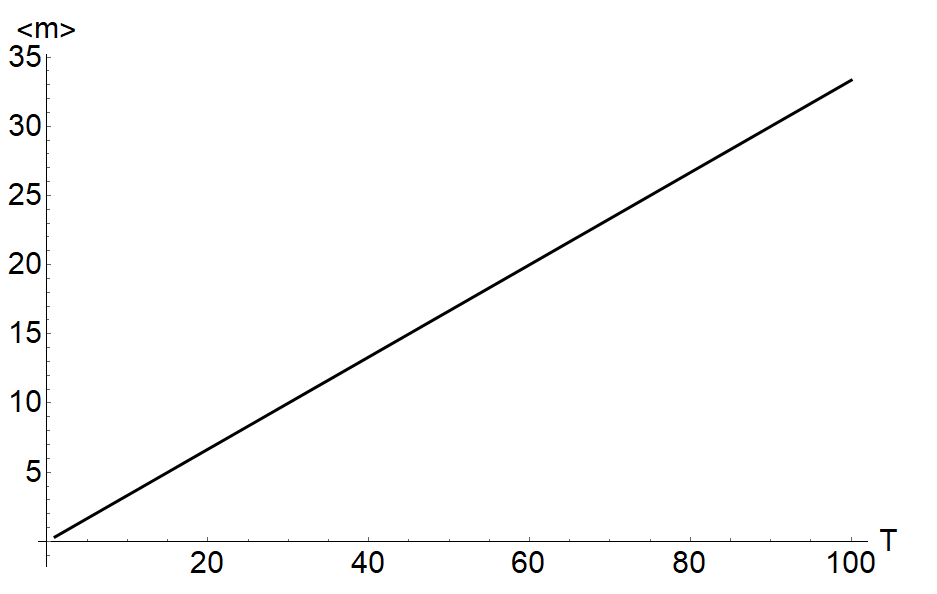}
\caption{Behavior of $\langle m \rangle$ according to the partition function $Z(T, \lambda)$, \eqref{eqn300}.}
\label{WA70}
\end{figure}
In Figure \ref{WA70}, the behavior of $\langle m \rangle$ is shown, and it is observed to exhibit a linear behavior $\langle m \rangle \propto T$, similar to that presented in Figure \ref{WA40}.

The heat capacity is obtained from \eqref{eqn241}
\begin{equation}
C=\frac{1}{3}
\label{eqn330}
\end{equation}

\begin{figure}[htbp]
\centering
		\includegraphics[width=0.5\textwidth]{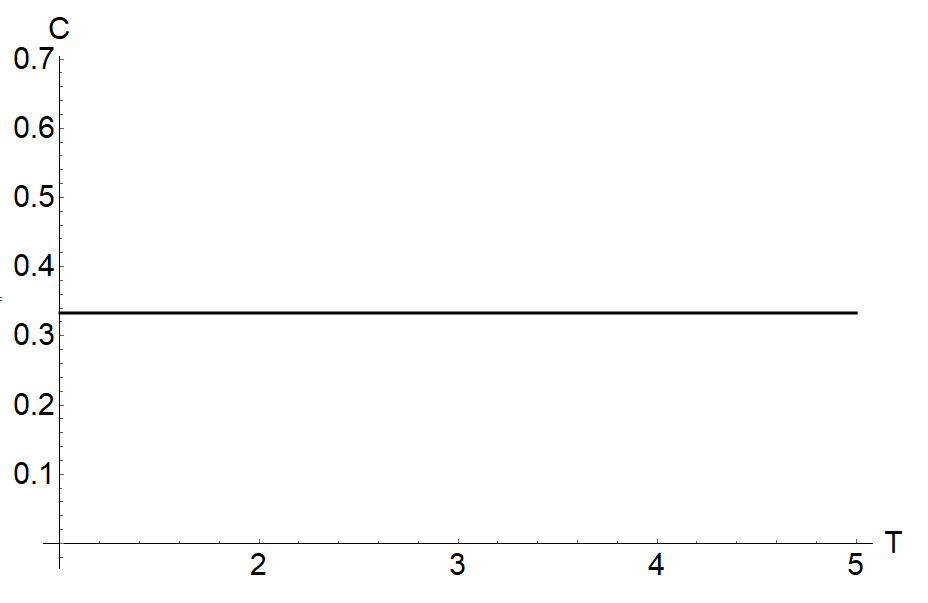}
\caption{Behavior of the heat capacity $C$ according to the partition function $Z(T, \lambda)$, \eqref{eqn300}.}
\label{WA80}
\end{figure}

In Figure \ref{WA80}, the behavior of $C$ is shown, and it is observed to be constant under the condition of cubic dominance. 

\subsection{A more realistic model for the CSDB.}

A more realistic description of the sector under study can include more macroeconomic characteristics. In this case, when including the Gross Domestic Product of the city of Bogot\'{a} and $m(\lambda)\approx \lambda^{3}$, allows us to write the partition function $Z(T,\bar{x})$ \eqref{eqn40}

\begin{align}
    	Z(T,\lambda,\lambda_{2})&=\int^{\infty}_{0}e^{-m(\lambda)}d\lambda \int^{x}_{0}f(\lambda_{2})d\lambda_{2} 
\notag\\
	&=\sqrt[3]{\frac{T}{\alpha}}\Gamma\left[\frac{4}{3}\right]P(x)
&\hspace{0.3cm}
\label{WA90}
\end{align}
 where 
\begin{equation}
P(x)=\frac{\alpha_{2} x^{4}}{4}+\frac{\beta_{2}x^{3}}{3}+\frac{\gamma_{2}x^{2}}{2}+\delta_{2}x.
\label{WA90a}
\end{equation}

The free money function, \eqref{eqn220}
\begin{equation}
f=-T\ln\left|\sqrt[3]{\frac{T}{\alpha}}\Gamma\left[\frac{4}{3}\right]P(x)\right|.
\label{WA100}
\end{equation}
The  entropy $S$, \eqref{eqn230}
\begin{equation}
S=\frac{\alpha}{3}+\ln\left|\sqrt[3]{\frac{T}{\alpha}}\Gamma\left[\frac{4}{3}\right]P(x)\right|.
\label{WA110}
\end{equation}
The average value of money per market agent $\left\langle m\right\rangle$, \eqref{eqn210}
\begin{equation}
\left\langle m\right\rangle=\frac{T\alpha}{3}.
\label{WA120}
\end{equation}
The term of work $y_{i}$, \eqref{eqn240}
\begin{equation}
y=\frac{T(\alpha_{2}x^{3}+\beta_{2}x^{2}+\gamma_{2}x+\delta_{2})}{P(x)}.
\label{WA130}
\end{equation}
And the heat capacity,\eqref{eqn241}
\begin{equation}
C=\frac{\alpha}{3}.
\label{WA140}
\end{equation}
\section{Geometrothermodynamics (GTD) for the sports sector in Bogot\'{a}} \label{sec6}
If the entropy $S$ is chosen as the thermodynamic potential \eqref{WA110}, then we have $\Phi = S$, $S = S(T, x)$, and the extensive variables $E^{a}=\left\{E^{1}, E^{2}\right\}=\left\{T,x\right\}$. Then
\begin{equation}
g^{I}=S\left[\frac{\partial^{2}S}{\partial T^{2}}+\frac{\partial^{2}S}{\partial x^{2}} \right],
\label{WA150}
\end{equation}
where
\begin{equation}
\frac{\partial^{2}S}{\partial x^{2}}=\frac{1}{3T^{2}},
\label{WA160}
\end{equation}
\begin{equation}
\frac{\partial^{2}S}{\partial x^{2}}=-\left[\frac{Q(x)^{2}}{P(x)^{2}}+\frac{Z(x)}{P(x)}\right],
\label{WA170}
\end{equation}
\begin{equation}
Q(x)=\alpha_{2}x^{3}+\beta_{2}x^{2}+\gamma_{2}x+\delta_{2},\,\,\,Q(x)=\frac{dP(x)}{dx}
\label{WA180}
\end{equation}
and
\begin{equation}
Z(x)=3\alpha_{2}x^{3}+2\beta_{2}x^{2}+\gamma_{2}.
\label{WA190}
\end{equation}
Therefore
\begin{equation}
g^{I}=S
\begin{pmatrix}
 \frac{1}{3T^{2}} & 0 \\
0 &  -\left[\frac{Q(x)^{2}}{P(x)^{2}}+\frac{Z(x)}{P(x)}\right] \\
\end{pmatrix},
\label{WA200}
\end{equation}
where \eqref{WA200} is the metric tensor $g^{I}$ defined on $\mathcal{E}$. The determinant $g^{I}$
\begin{equation}
\mbox{det}g^{I}=\frac{S}{3T^{2}}\left[\frac{Q(x)^{2}}{P(x)^{2}}+\frac{Z(x)}{P(x)}\right],
\label{WA210}
\end{equation}
this vanishes under two conditions
\begin{equation}
\frac{S}{3T^{2}}=0
\label{WA220}
\end{equation}
which implies zero entropy $S=0$, which is not possible, also
\begin{equation}
\frac{Q(x)^{2}}{P(x)}=-Z(x).
\label{WA230}
\end{equation}
In Geometrothermodynamics, the determinant of the metric, $\mbox{det}(g)$, is a scalar that characterizes the geometry of the thermodynamic state space. If the determinant of the metric $\mbox{det}(g)$ is zero, it implies that the metric is degenerate, meaning that the thermodynamic state space has a singular geometric structure at that point.  Physically, a zero determinant can indicate: a critical point, a bifurcation point where the system's behavior qualitatively changes, such as a shift in stability, a thermodynamic singularity where the system's thermodynamic properties are not well-defined.

On the other hand, a direct calculation of $g_{II}$ contained in \eqref{gtd30} allows us to obtain
\begin{equation}
g^{II}=S\left[-\frac{\partial^{2}S}{\partial T^{2}}+\frac{\partial^{2}S}{\partial x^{2}} \right],
\label{WA240}
\end{equation}
so
\begin{equation}
g^{II}=S
\begin{pmatrix}
 \frac{1}{3T^{2}} & 0 \\
0 &  -\frac{Q(x)^{2}}{P(x)^{2}}+\frac{Z(x)}{P(x)} \\
\end{pmatrix}.
\label{WA250}
\end{equation}
Also, from \eqref{gtd40}
\begin{equation}
g^{III}=T\frac{\partial S}{\partial T}\frac{\partial^{2} S}{\partial T^{2}}dT^{2}+\left[T\frac{\partial S}{\partial T}+x\frac{\partial S}{\partial x}\right]\frac{\partial^{2} S}{\partial T\partial x}dTdx+x\frac{\partial S}{\partial x}\frac{\partial^{2} S}{\partial x^{2}}dx^{
2}
\label{WA260},
\end{equation}
thus
\begin{equation}
g^{III}=S
\begin{pmatrix}
 \frac{1}{9T^{2}} & 0 \\
0 & \frac{xQ(x)}{P(x)} \left[\frac{Q(x)^{2}}{P(x)^{2}}+\frac{Z(x)}{P(x)}\right] \\
\end{pmatrix}.
\label{WA270}
\end{equation}
From \eqref{WA200}, \eqref{WA250} and \eqref{WA270}, it is possible to calculate the Kretschmann scalars   $K_{I}$, $K_{II}$ and $K_{III}$  and the Ricci scalars $R_{I}$, $R_{II}$ and $R_{III}$ for the metrics  $g_{I}$, $g_{II}$ and $g_{III}$ which are explicitly shown in Appendices \ref{apend000}, \ref{apend010}, \ref{apend020}, and \ref{apend030}.

\begin{figure}[htbp]
\centering
		\includegraphics[width=0.7\textwidth]{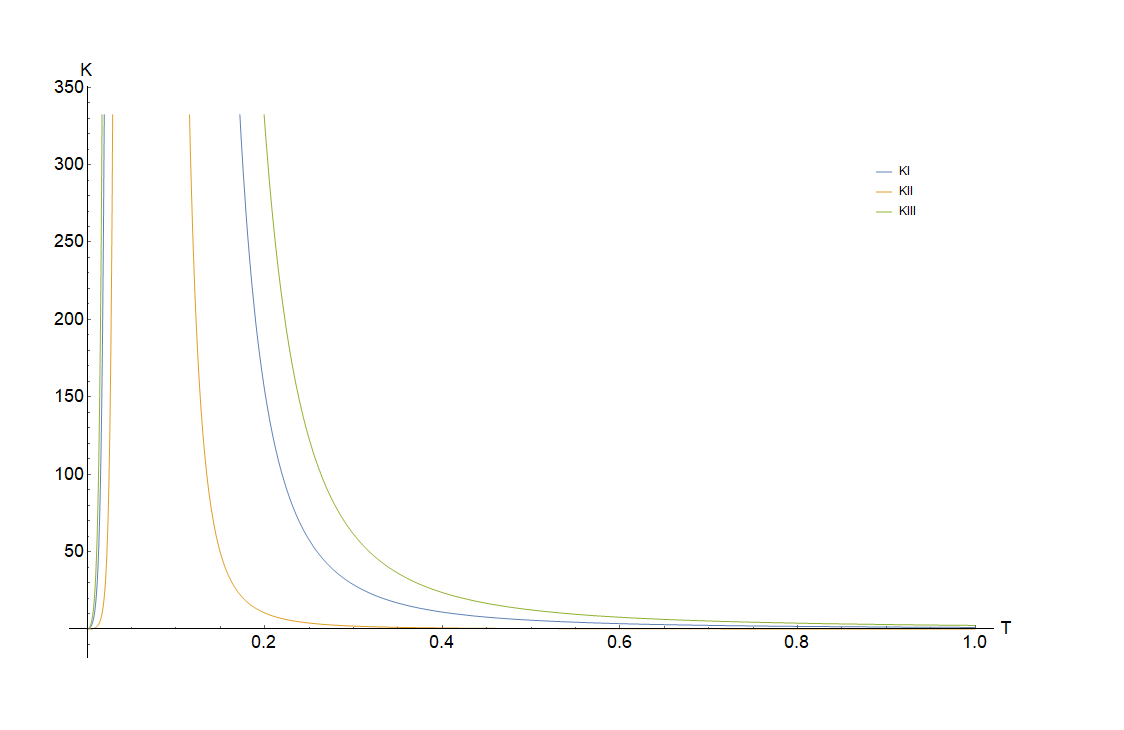}
\caption{Behavior of  $K_{I}$, $K_{II}$ and $K_{III}$ as a function of temperature  $T$ with $x=cte$. }
\label{Ksc0}
\end{figure}

\begin{figure}[htbp]
\centering
		\includegraphics[width=0.7\textwidth]{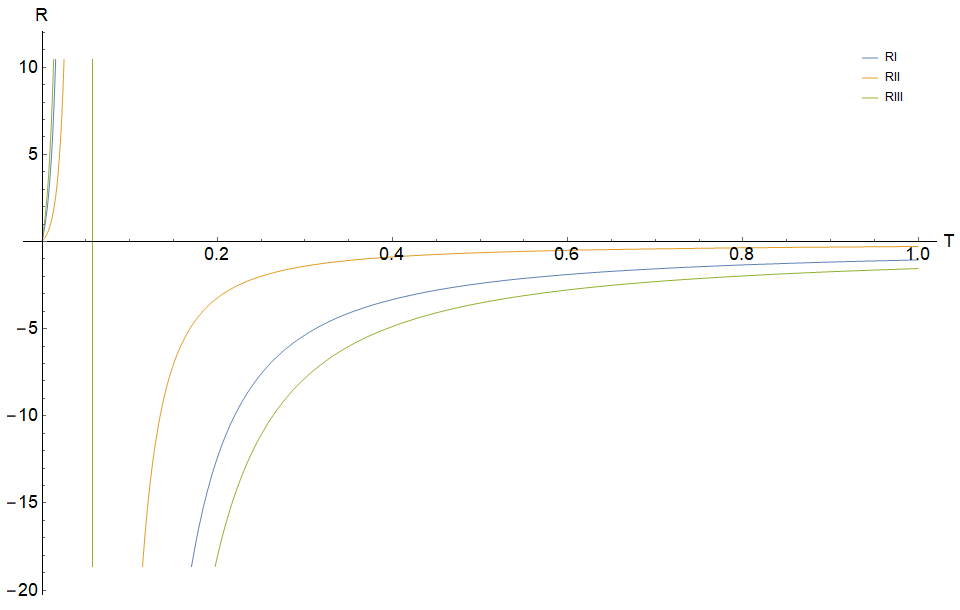}
\caption{Behavior of $R_{I}$, $R_{II}$, $R_{III}$ as a function of temperature $T$ with $x=cte$. }
\label{Ksc10}
\end{figure}

In Figure \ref{Ksc0}, the behavior of $K_{I}$, $K_{II}$, and $K_{III}$ as a function of temperature $T$ with $x$ held constant is shown. It is observed that $K_{I}$, $K_{II}$, and $K_{III}$ exhibit a singularity at $T \approx 0.1$, which can be interpreted as an economic crisis. This point is sensitive, as the analyzed period for the sector corresponds to 2018-2022, which includes the COVID-19 pandemic period, declared by the World Health Organization on January 30, 2020, and lasting until May 5, 2023 \cite{wikipedia2024pandemia}.

Due to the presence of the virus, many of the collective and outdoor activities included in the CSDB had to be restricted. Another notable characteristic is that in the $\lim_{T\longrightarrow \infty} \left\{K_{i}, R_{i}\right\}\approx 0$, this means that the curvature of the $K_{i}$ and $R_{i}$ diminishes over time, which is interpreted as low thermodynamic interaction in relation to Bogot\'{a}'s GDP \cite{Quevedo2008xn}.

Identical conclusions can be drawn from Figure \ref{Ksc10}, which shows the behavior of $R_{i}$ as a function of temperature $T$ with $x=cte$. It is observed that there is a singularity at $T\approx 0.1$,  whose origin is associated with the pandemic period. An interesting property of this work is tha
\begin{equation}
K_{i}\propto -R_{i}.
\label{WA280}
\end{equation}
\begin{figure}[htbp]
\centering
		\includegraphics[width=0.7\textwidth]{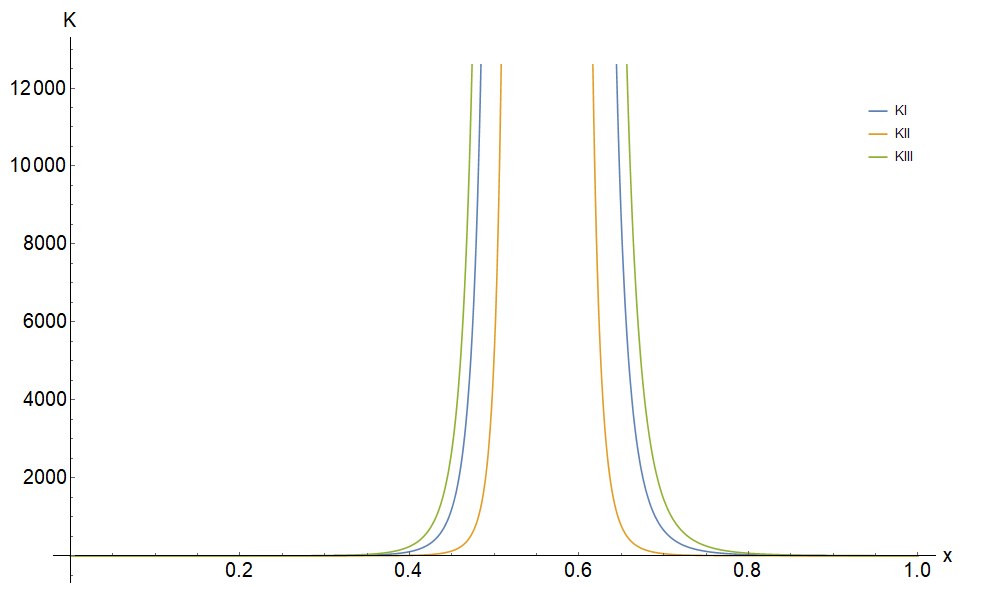}
\caption{Behavior of  $K_{I}$, $K_{II}$ and $K_{III}$ as a function of $x$  with $T=cte$. }
\label{Ksc20}
\end{figure}
\begin{figure}[htbp]
\centering
		\includegraphics[width=0.7\textwidth]{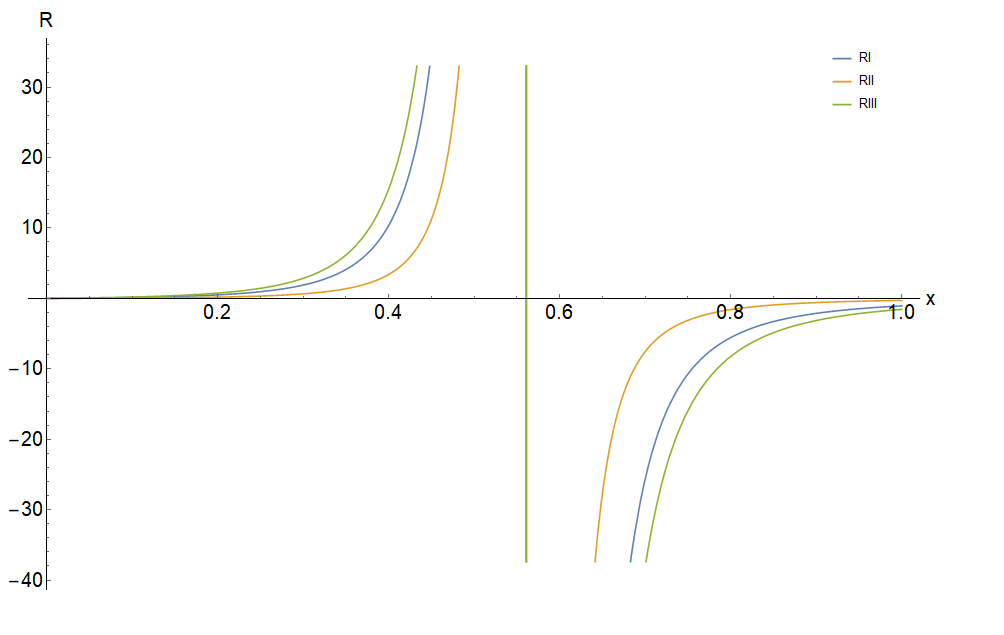}
\caption{Behavior of  $R_{I}$, $R_{II}$ and $R_{III}$ as a function of $x$  with $T=cte$. }
\label{Ksc30}
\end{figure}

In Figure \ref{Ksc20}, the behavior $K_{I}$, $K_{II}$ y $K_{III}$  as a function of $x$ with $T=cte$  is shown. It is observed that there is a singularity at $x\approx 0.5$, which can be explained by the aforementioned reasons. Similarly, Figure \ref{Ksc30} shows the behavior of $R_{I}$, $R_{II}$, $R_{III}$ as a function of $x$ with $T=cte$.  Here $K_{i}$ and $R_{i}$ exhibit singularities for the same values. In Appendices \ref{apend000}, \ref{apend010}, \ref{apend020}, and \ref{apend030}, the functional forms of the determinant of $g^{i}$ and the curvature scalars $K_{i}$ and $R_{i}$ are presented, highlighting the characteristics that are common among them.

\section{Discussions and Conclusions} \label{sec7}
The present work presents a detailed analysis of the CSDB in Bogot\'{a} based on a thermoeconomic model, as mentioned in \cite{Quevedo2011}. In this context, income elasticity $\lambda$  is employed as an essential microeconomic parameter to identify whether sports are considered a normal or inferior good. The results show that, in certain years, the elasticity is positive and greater than 1, indicating that sports are perceived as a normal or even luxury good \cite{dane2023}.

One of the key findings of this study is that spending on sports in Bogot\'{a} is highly sensitive to changes in GDP, suggesting that as the local economy improves, the supply of sports services also increases significantly.

This behavior supports the idea that sports services are seen as normal or luxury goods. Conversely, when the elasticity is less than 1, it suggests that spending on sports is relatively inelastic with respect to GDP, indicating that sports services are considered a basic necessity or face structural barriers that limit the supply response to economic changes.

The CSDB provides detailed information on the behavior of various economic sectors participating in the analysis, such as manufacturing, apparel, and services. According to the model, the average value of money among agents depends on how commercial transactions are conducted, which are modeled as elastic collisions between two particles in one dimension. In these transactions, money is exchanged for goods and services related to sports, reflecting a linear relationship between the average value of money  $ \langle m \rangle $  and economic temperature $T$, as observed in Figures \ref{WA40} and \ref{WA70}.

In the analysis presented, if the partition function has a cubic domain, $ m(\lambda) \propto \lambda^3 $, the entropy follows the Boltzmann Principle  meaning that there exists a set of microstates compatible with a given macrostate \cite{alonso1976fisica,reif1996fisica,landau1988fisica,greiner2012thermodynamics}. This result reinforces the connection between the economic model and the laws of thermodynamics.

Another important aspect addressed is the relationship between economic temperature and the average value of money per agent. In the cases studied, it is shown that there is a linear dependency between these two parameters, suggesting that economic temperature is a good indicator of economic activity in the sports sector. Additionally, when the partition function has a quadratic domain, $m(\lambda)\propto \lambda^2 $, it is observed that the heat capacity of the CSDB decreases as temperature increases, as shown in Figure \ref{WA50}. This behavior may indicate a decrease in the sector's capacity to absorb more money, equating to a lower capacity for investment.

Geometrothermodynamics applied to the CSDB reveals that curvature scalars, such as the Kretschmann and Ricci scalars, show singularities that are interpreted as economic crises during the period studied.

This result is consistent with the crisis provoked by the COVID-19 pandemic, suggesting that the model adequately captures fluctuations in the economic stability of the sports sector. These geometric metrics provide a profound interpretation of the system's stability, allowing for an analysis of the sector's robustness against disruptive events.

The study demonstrates the application of geometrothermodynamics (GTD) in econophysics for describing the behavior of economic systems. 
The results suggest that the sports sector responds elastically to changes in GDP, indicating that it is perceived as a normal or luxury good. However, the decrease in heat capacity with the increase in economic temperature suggests that there are limitations on the future growth of the sector. This trend raises questions about the ability of the sports sector to absorb more money in the long term. Future research could focus on developing additional metrics to measure the sensitivity of different sectors to changes in the economic environment
, incorporating elements such as economic entropy and heat capacity as indicators of stability. Moreover, it is recommended to continue exploring more complex models that include multidimensional dynamics and the interaction of multiple macroeconomic factors. This could provide a more comprehensive view of how economic sectors evolve during times of crisis and recovery.

\appendix
\section{Determinant of $g_{I}$} \label{apend000}
Starting from \eqref{WA90}, \eqref{WA110}, \eqref{WA180}, \eqref{WA190}, and \eqref{WA200}, the determinant of 
$g_{I}$ can be explicitly calculated as
\begin{equation}
\mbox{det}g = \frac{4NL^2}{9T^2x^2D^2},
\label{Aeq00}
\end{equation}
where
\begin{align*}
N &= (12\delta_2^2 + 12x(3\gamma_2 + x(4\beta_2 + 5\alpha_2x)))\delta_2 \\
  &\quad + x^2(18\gamma_2^2 + 5x(8\beta_2 + 9\alpha_2x)\gamma_2 \\
  &\quad + x^2(20\beta_2^2 + 42\alpha_2x\beta_2 + 21\alpha_2^2x^2)) \\[2ex]
D &= 12\delta_2 + x(6\gamma_2 + x(4\beta_2 + 3\alpha_2x)) \\[2ex]
L &= \alpha_2 + 3\log\left(\frac{1}{12}\sqrt[3]{\frac{T}{\alpha}}x D \Gamma \left[\frac{4}{3}\right]\right),
\end{align*}
Thus, for \eqref{Aeq00}, $N$ and $D$  are functions dependent on the functional form of the time series of income elasticity  $m(\lambda)$, \eqref{aqn1550}; GDP,  \eqref{WA11}, and $L$ is associated with the entropy of the system $S$, \eqref{WA110}.

\newpage
\section{Curvature Scalars}\label{apend010}
\begin{equation}
K_{I} = \frac{81N^2}{D^4L^6}
\label{Aeq10}
\end{equation}
 Where
\begin{align*}
N &= 576\delta_2^4 + 72x(({\alpha_2}+42)\gamma_2 + x(4({\alpha_2}+13)\beta_2 + {\alpha_2}(9{\alpha_2}+62)x))\delta_2^3 \\
  &\quad + 6x^2(6({\alpha_2}+150)\gamma_2^2 + x(8(5{\alpha_2}+266)\beta_2 + 9{\alpha_2}(13{\alpha_2}+274)x)\gamma_2 \\
  &\quad + x^2(1232\beta_2^2 + 6{\alpha_2}({\alpha_2}+470)x\beta_2 + 9{\alpha_2}^2(178-3{\alpha_2})x^2))\delta_2^2 \\
  &\quad + x^3(3888\gamma_2^3 + 12x((5{\alpha_2}+1092)\beta_2 + 24{\alpha_2}({\alpha_2}+51)x)\gamma_2^2 \\
  &\quad + 12x^2(-2({\alpha_2}-598)\beta_2^2 + 2{\alpha_2}(2{\alpha_2}+1319)x\beta_2 - 9({\alpha_2}-160){\alpha_2}^2x^2)\gamma_2 \\
  &\quad + x^3(-32({\alpha_2}-159)\beta_2^3 - 96({\alpha_2}-172){\alpha_2}x\beta_2^2 \\
  &\quad - 9{\alpha_2}^2(19{\alpha_2}-1960)x^2\beta_2 + 9{\alpha_2}^3(688-9{\alpha_2})x^3))\delta_2 \\
  &\quad + x^4(1197{\alpha_2}^4x^8 + 4788{\alpha_2}^3\beta_2x^7 \\
  &\quad + 3{\alpha_2}^2(({\alpha_2}+2368)\beta_2^2 - 3({\alpha_2}-558){\alpha_2}\gamma_2)x^6 \\
  &\quad + {\alpha_2}\beta_2(4({\alpha_2}+1158)\beta_2^2 - 3({\alpha_2}-4892){\alpha_2}\gamma_2)x^5 \\
  &\quad + (1120\beta_2^4 + 6{\alpha_2}(3{\alpha_2}+2354)\gamma_2\beta_2^2 + 9{\alpha_2}^2({\alpha_2}+825)\gamma_2^2)x^4 \\
  &\quad + 20\beta_2\gamma_2(224\beta_2^2 + 3{\alpha_2}({\alpha_2}+234)\gamma_2)x^3 \\
  &\quad + 2\gamma_2^2(4({\alpha_2}+821)\beta_2^2 + 27{\alpha_2}({\alpha_2}+84)\gamma_2)x^2 \\
  &\quad + 12({\alpha_2}+348)\beta_2\gamma_2^3x + 972\gamma_2^4) \\
  &\quad + 3x({\alpha_2}(\beta_2^2-3{\alpha_2}\gamma_2)x^6 + 3{\alpha_2}(\beta_2\gamma_2-9{\alpha_2}\delta_2)x^5 \\
  &\quad + 3{\alpha_2}(3\gamma_2^2-7\beta_2\delta_2)x^4 + 2(-4\delta_2\beta_2^2+\gamma_2^2\beta_2+15{\alpha_2}\delta_2\gamma_2)x^3 \\
  &\quad + 6\delta_2(9{\alpha_2}\delta_2+\beta_2\gamma_2)x^2 + 24\beta_2\delta_2^2x + 6\delta_2^2\gamma_2) \\
  &\quad \times (12\delta_2+x(6\gamma_2+x(4\beta_2+3{\alpha_2}x))) \\
  &\quad \times \log\left(\frac{1}{12}\sqrt[3]{\frac{T}{{\alpha_2}}}x(12\delta_2+x(6\gamma_2+x(4\beta_2+3{\alpha_2}x)))\Gamma\left[\frac{4}{3}\right]\right) \\
D &= 12\delta_2^2 + 12x(3\gamma_2+x(4\beta_2+5{\alpha_2}x))\delta_2 \\
  &\quad + x^2(18\gamma_2^2 + 5x(8\beta_2+9{\alpha_2}x)\gamma_2 + x^2(20\beta_2^2+42{\alpha_2}x\beta_2+21{\alpha_2}^2x^2)) \\
L &= {\alpha} + 3\log\left(\frac{1}{12}\sqrt[3]{\frac{T}{{\alpha}}}x(12\delta_2+x(6\gamma_2+x(4\beta_2+3{\alpha_2}x)))\Gamma\left[\frac{4}{3}\right]\right)
\end{align*}
\begin{equation}
R_{I} = -\frac{9N}{D^2L^3}
\label{Aeq20}
\end{equation}
where $N$, $D$, and $L$ are the same as in the previous expression for $K_{I}$.
\begin{equation}
K_{II} = \frac{81N^2}{D^4L^6}
\label{Aeq30}
\end{equation}
where
\begin{align*}
N &= 288\delta_2^4 + 72x(({\alpha_2}+18)\gamma_2 + x(4({\alpha_2}+5)\beta_2 + {\alpha_2}(9{\alpha_2}+22)x))\delta_2^3 \\
  &\quad + 6x^2(6({\alpha_2}+54)\gamma_2^2 + x(8(5{\alpha_2}+82)\beta_2 + 9{\alpha_2}(13{\alpha_2}+74)x)\gamma_2 \\
  &\quad + x^2(304\beta_2^2 + 6{\alpha_2}({\alpha_2}+94)x\beta_2 + 9{\alpha_2}^2(26-3{\alpha_2})x^2))\delta_2^2 \\
  &\quad + x^3(-9{\alpha_2}^3(9{\alpha_2}-128)x^6 - 9{\alpha_2}^2(19{\alpha_2}-392)\beta_2x^5 \\
  &\quad - 12{\alpha_2}(8({\alpha_2}-38)\beta_2^2 + 9({\alpha_2}-32){\alpha_2}\gamma_2)x^4 \\
  &\quad - 8\beta_2(4({\alpha_2}-39)\beta_2^2 - 3{\alpha_2}(2{\alpha_2}+307)\gamma_2)x^3 \\
  &\quad + 24\gamma_2(6{\alpha_2}(2{\alpha_2}+27)\gamma_2 - ({\alpha_2}-158)\beta_2^2)x^2 \\
  &\quad + 12(5{\alpha_2}+324)\beta_2\gamma_2^2x + 1296\gamma_2^3)\delta_2 \\
  &\quad + x^4(315{\alpha_2}^4x^8 + 1260{\alpha_2}^3\beta_2x^7 \\
  &\quad + 3{\alpha_2}^2(({\alpha_2}+632)\beta_2^2 - 3({\alpha_2}-138){\alpha_2}\gamma_2)x^6 \\
  &\quad + {\alpha_2}\beta_2(4({\alpha_2}+318)\beta_2^2 - 3({\alpha_2}-1252){\alpha_2}\gamma_2)x^5 \\
  &\quad + (320\beta_2^4 + 6{\alpha_2}(3{\alpha_2}+634)\gamma_2\beta_2^2 + 9{\alpha_2}^2({\alpha_2}+207)\gamma_2^2)x^4 \\
  &\quad + 4\beta_2\gamma_2(320\beta_2^2 + 3{\alpha_2}(5{\alpha_2}+318)\gamma_2)x^3 \\
  &\quad + 2\gamma_2^2(4({\alpha_2}+241)\beta_2^2 + 27{\alpha_2}({\alpha_2}+24)\gamma_2)x^2 \\
  &\quad + 12({\alpha_2}+108)\beta_2\gamma_2^3x + 324\gamma_2^4) \\
  &\quad + 3x({\alpha_2}(\beta_2^2-3{\alpha_2}\gamma_2)x^6 + 3{\alpha_2}(\beta_2\gamma_2-9{\alpha_2}\delta_2)x^5 \\
  &\quad + 3{\alpha_2}(3\gamma_2^2-7\beta_2\delta_2)x^4 + 2(-4\delta_2\beta_2^2+\gamma_2^2\beta_2+15{\alpha_2}\delta_2\gamma_2)x^3 \\
  &\quad + 6\delta_2(9{\alpha_2}\delta_2+\beta_2\gamma_2)x^2 + 24\beta_2\delta_2^2x + 6\delta_2^2\gamma_2) \\
  &\quad \times (12\delta_2+x(6\gamma_2+x(4\beta_2+3{\alpha_2}x))) \\
  &\quad \times \log\left(\frac{1}{12}\sqrt[3]{\frac{T}{{\alpha}}}x(12\delta_2+x(6\gamma_2+x(4\beta_2+3{\alpha_2}x)))\Gamma\left[\frac{4}{3}\right]\right)
\end{align*}
$D$ and $L$ are the same as in the previous expression for $K_{I}$.
\begin{equation}
R_{II} = -\frac{9N}{D^2L^3}
\label{Aeq40}
\end{equation}
where $N$, $D$ and $L$ are the same as in the previous expression for $K_{II}$.
\begin{equation}
K_{III} = \frac{9N^2}{64D^4L^6}
\label{Aeq50}
\end{equation}
so
\begin{align*}
N &= 20736\delta_2^5 + 144x(({\alpha_2}+50)\gamma_2 + x(16(4{\alpha_2}+69)\beta_2 + 9{\alpha_2}(15{\alpha_2}+146)x))\delta_2^4 \\
  &\quad + 12x^2(36(11{\alpha_2}+696)\gamma_2^2 + 24x(4(19{\alpha_2}+618)\beta_2 + 9{\alpha_2}(17{\alpha_2}+320)x)\gamma_2 \\
  &\quad + x^2(256(4{\alpha_2}+135)\beta_2^2 + 96{\alpha_2}(32{\alpha_2}+831)x\beta_2 + 27{\alpha_2}^2(63{\alpha_2}+1696)x^2))\delta_2^3 \\
  &\quad + 12x^3(27{\alpha_2}^3(27{\alpha_2}+1846)x^6 + 9{\alpha_2}^2(245{\alpha_2}+15276)\beta_2x^5 \\
  &\quad + 3{\alpha_2}(8(85{\alpha_2}+5214)\beta_2^2 + 27{\alpha_2}(35{\alpha_2}+1598)\gamma_2)x^4 \\
  &\quad + 16\beta_2(16(2{\alpha_2}+147)\beta_2^2 + 9{\alpha_2}(35{\alpha_2}+1611)\gamma_2)x^3 \\
  &\quad + 16\gamma_2((103{\alpha_2}+6450)\beta_2^2 + 27{\alpha_2}(7{\alpha_2}+243)\gamma_2)x^2 \\
  &\quad + 12(125{\alpha_2}+7692)\beta_2\gamma_2^2x + 36(7{\alpha_2}+750)\gamma_2^3)\delta_2^2 \\
  &\quad + x^4(81{\alpha_2}^4(15{\alpha_2}+3128)x^8 + 36{\alpha_2}^3(151{\alpha_2}+27276)\beta_2x^7 \\
  &\quad + 27{\alpha_2}^2(4(79{\alpha_2}+13072)\beta_2^2 + 3{\alpha_2}(101{\alpha_2}+12304)\gamma_2)x^6 \\
  &\quad + 24{\alpha_2}\beta_2(16(14{\alpha_2}+2325)\beta_2^2 + 9{\alpha_2}(115{\alpha_2}+13062)\gamma_2)x^5 \\
  &\quad + 2(128(4{\alpha_2}+819)\beta_2^4 + 48{\alpha_2}(235{\alpha_2}+27456)\gamma_2\beta_2^2 + 243{\alpha_2}^2(35{\alpha_2}+2836)\gamma_2^2)x^4 \\
  &\quad + 16\beta_2\gamma_2(8(41{\alpha_2}+6354)\beta_2^2 + 9{\alpha_2}(205{\alpha_2}+17616)\gamma_2)x^3 \\
  &\quad + 12\gamma_2^2(20(37{\alpha_2}+4824)\beta_2^2 + 9{\alpha_2}(113{\alpha_2}+7368)\gamma_2)x^2 \\
  &\quad + 288(19{\alpha_2}+2490)\beta_2\gamma_2^3x + 648({\alpha_2}+252)\gamma_2^4)\delta_2 \\
  &\quad + x^5(36288{\alpha_2}^5x^{10} + 63{\alpha_2}^4({\alpha_2}+2904)\beta_2x^9 \\
  &\quad + 12{\alpha_2}^3((19{\alpha_2}+30474)\beta_2^2 + 27{\alpha_2}({\alpha_2}+604)\gamma_2)x^8 \\
  &\quad + 3{\alpha_2}^2\beta_2(4(23{\alpha_2}+30180)\beta_2^2 + 3{\alpha_2}(173{\alpha_2}+85872)\gamma_2)x^7 \\
  &\quad + 4{\alpha_2}(4(7{\alpha_2}+11106)\beta_2^4 + 15{\alpha_2}(41{\alpha_2}+18894)\gamma_2\beta_2^2 + 54{\alpha_2}^2(7{\alpha_2}+1860)\gamma_2^2)x^6 \\
  &\quad + 2\beta_2(17280\beta_2^4 + 32{\alpha_2}(22{\alpha_2}+11439)\gamma_2\beta_2^2 + 9{\alpha_2}^2(253{\alpha_2}+64596)\gamma_2^2)x^5 \\
  &\quad + 8\gamma_2(20({\alpha_2}+1098)\beta_2^4 + 24{\alpha_2}(20{\alpha_2}+5787)\gamma_2\beta_2^2 + 27{\alpha_2}^2(11{\alpha_2}+1806)\gamma_2^2)x^4 \\
  &\quad + 12\beta_2\gamma_2^2(4(13{\alpha_2}+7308)\beta_2^2 + 3{\alpha_2}(107{\alpha_2}+20424)\gamma_2)x^3 \\
  &\quad + 48\gamma_2^3((17{\alpha_2}+7158)\beta_2^2 + 27{\alpha_2}({\alpha_2}+138)\gamma_2)x^2 \\
  &\quad + 72(5{\alpha_2}+2292)\beta_2\gamma_2^4x + 31104\gamma_2^5) \\
  &\quad + 3x(21{\alpha_2}^3\beta_2x^{10} + 12{\alpha_2}^2(4\beta_2^2 + 9{\alpha_2}\gamma_2)x^9 \\
  &\quad + {\alpha_2}(28\beta_2^3 + 333{\alpha_2}\gamma_2\beta_2 + 405{\alpha_2}^2\delta_2)x^8 \\
  &\quad + 4{\alpha_2}(70\gamma_2\beta_2^2 + 9{\alpha_2}(8\gamma_2^2 + 33\beta_2\delta_2))x^7 \\
  &\quad + (40\gamma_2\beta_2^3 + 12{\alpha_2}(39\gamma_2^2 + 89\beta_2\delta_2)\beta_2 + 1485{\alpha_2}^2\delta_2\gamma_2)x^6 \\
  &\quad + 8(32\delta_2\beta_2^3 + 12\gamma_2^2\beta_2^2 + 324{\alpha_2}\delta_2\gamma_2\beta_2 + 27{\alpha_2}(\gamma_2^3 + 6{\alpha_2}\delta_2^2))x^5 \\
  &\quad + 4(\beta_2\gamma_2(15\gamma_2^2 + 202\beta_2\delta_2) + 9{\alpha_2}\delta_2(43\gamma_2^2 + 65\beta_2\delta_2))x^4 \\
  &\quad + 24\delta_2(32\delta_2\beta_2^2 + 30\gamma_2^2\beta_2 + 117{\alpha_2}\delta_2\gamma_2)x^3 \\
  &\quad + 36\delta_2(3\gamma_2^3 + 38\beta_2\delta_2\gamma_2 + 45{\alpha_2}\delta_2^2)x^2 \\
  &\quad + 96\delta_2^2(3\gamma_2^2 + 8\beta_2\delta_2)x + 216\delta_2^3\gamma_2) \\
  &\quad \times (12\delta_2 + x(6\gamma_2 + x(4\beta_2 + 3{\alpha_2}x))) \\
  &\quad \times \log\left(\frac{1}{12}\sqrt[3]{\frac{T}{{\alpha}}}x(12\delta_2 + x(6\gamma_2 + x(4\beta_2 + 3{\alpha_2}x)))\Gamma\left[\frac{4}{3}\right]\right) \\
D &= \delta_2 + x(\gamma_2 + x(\beta_2 + {\alpha_2}x)) \\
L &= {\alpha} + 3\log\left(\frac{1}{12}\sqrt[3]{\frac{T}{{\alpha}}}x(12\delta_2 + x(6\gamma_2 + x(4\beta_2 + 3{\alpha_2}x)))\Gamma\left[\frac{4}{3}\right]\right)
\end{align*}
\begin{equation}
R_{III} = -\frac{3N}{8D^2L^3}
\label{Aeq60}
\end{equation}
$N$ are the same as in the previous expression for $K_{III}$, y:
\begin{align*}
D &= 12\delta_2^2 + 12x(3\gamma_2 + x(4\beta_2 + 5{\alpha_2}x))\delta_2 \\
  &\quad + x^2(18\gamma_2^2 + 5x(8\beta_2 + 9{\alpha_2}x)\gamma_2 + x^2(20\beta_2^2 + 42{\alpha_2}x\beta_2 + 21{\alpha_2}^2x^2)) \\
L &= {\alpha} + 3\log\left(\frac{1}{12}\sqrt[3]{\frac{T}{{\alpha}}}x(12\delta_2 + x(6\gamma_2 + x(4\beta_2 + 3{\alpha_2}x)))\Gamma\left[\frac{4}{3}\right]\right)
\end{align*}
\section{Relationship between the denominators of the scalars and the determinant of $ g_I $}\label{apend020}
The denominator in the expressions of the scalars and the determinant of $ g_I $ has a similar structure, which involves the functions $ D $ and $ L $. The corresponding formulas are described below:
\begin{itemize}
	\item  \textbf{Determinant of $ g_I $}:
   \begin{equation}
   \text{det}(g) = \frac{4 N L^2}{9 T^2 x^2 D^2}
   \end{equation}
  Here, the denominators include $ T^2 x^2 D^2 $, where $ D $ is a  function of $ \delta_2 $, $ \gamma_2 $, $ \beta_2 $ y $ \alpha_2 $
	
	\item \textbf{Scalars $ K_I $, $ K_{II} $, $ K_{III} $}:
   \begin{equation}
   K = \frac{81 N^2}{D^4 L^6}
   \end{equation}
   In these scalars, the denominators include $ D^4 L^6 $, which indicates that these functions depend on both $ D $ and $ L $, similar to the determinant of $ g $.
	
	\item \textbf{Scalars $ R_I $, $ R_{II} $, $ R_{III} $}:
   \begin{equation}
   R = -\frac{9 N}{D^2 L^3}
   \end{equation}
  Here, the denominator is $ D^2 L^3 $, which means that the same functions$ D $ and $ L $ govern the geometric scale of the scalars.
\end{itemize}
\subsection{Relationship between the denominators}
\begin{itemize}
	\item \textbf{Function $ D $}: It appears as  $ D^2 $  in the determinant and as $ D^4 $  in the curvature scalars. This reflects the relationship between the geometric properties of the metric and the associated curvatures, where $ D $  represents a combination of the variables  $ \delta_2 $, $ \gamma_2 $, $ \beta_2 $, $ \alpha_2 $ in terms of $ x $.
	 \item \textbf{Function $ L $}: It is present in the curvature scalars but not in the determinant. The function  $ L $ appears raised to the sixth power in the curvature scalars, indicating its role in the normalization of the curvatures, while the determinant depends only on the square of $ L $.
\end{itemize}

\clearpage

\section{Surface diagrams for the curvature scalars}\label{apend030}

\begin{figure}[htbp]
\centering
		\includegraphics[width=0.7\textwidth]{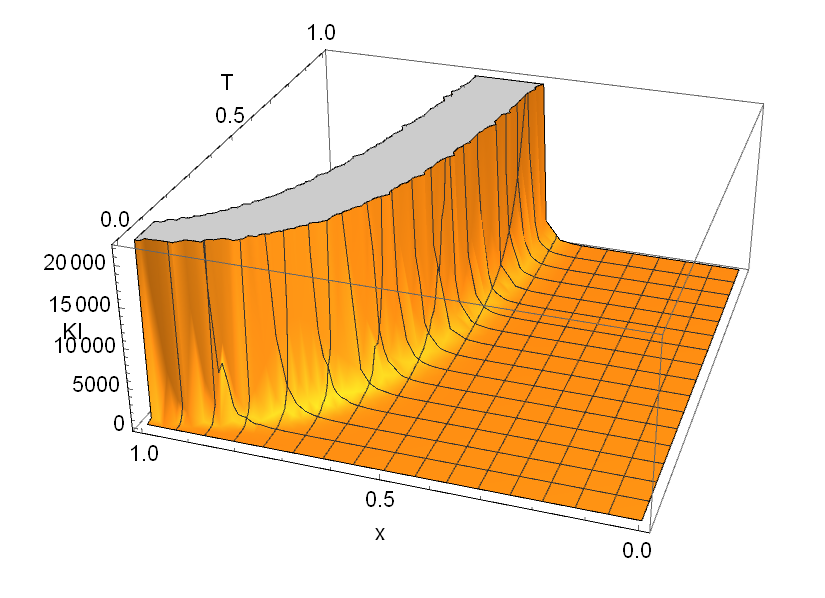}
\caption{Surface diagram $K_{I}$.}
\label{QG0}
\end{figure}
\begin{figure}[htbp]
\centering
		\includegraphics[width=0.7\textwidth]{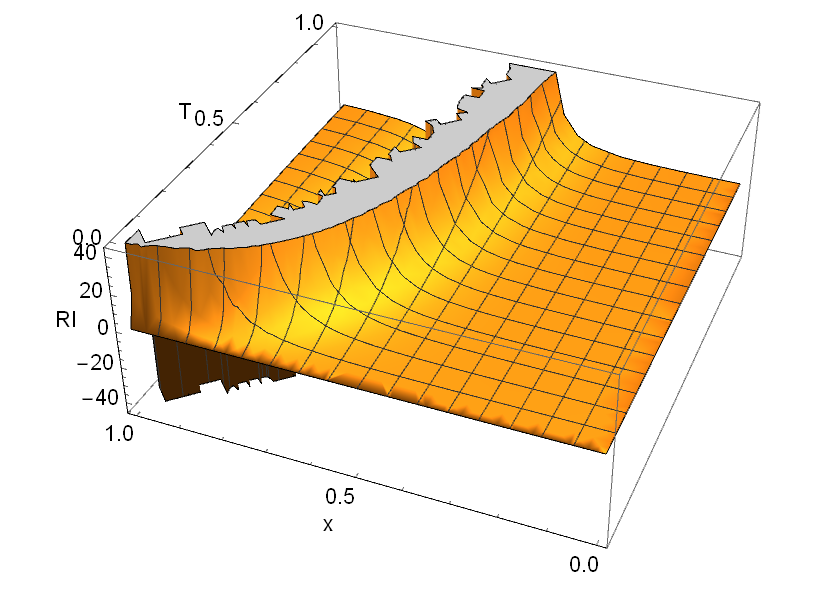}
\caption{Surface diagram $R_{I}$.}
\label{QG10}
\end{figure}

\begin{figure}[htbp]
\centering
		\includegraphics[width=0.7\textwidth]{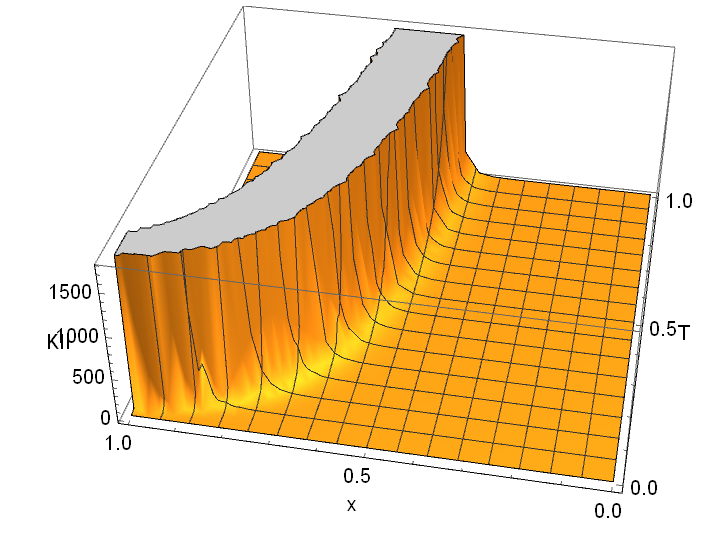}
\caption{Surface diagram $K_{II}$.}
\label{QG20}
\end{figure}
\begin{figure}[htbp]
\centering
		\includegraphics[width=0.7\textwidth]{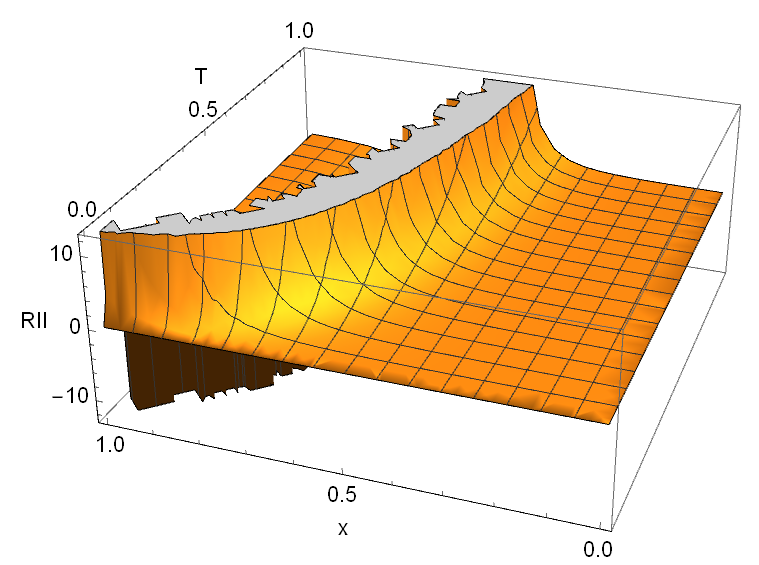}
\caption{Surface diagram $R_{II}$.}
\label{QG30}
\end{figure}
\begin{figure}[htbp]
\centering
		\includegraphics[width=0.7\textwidth]{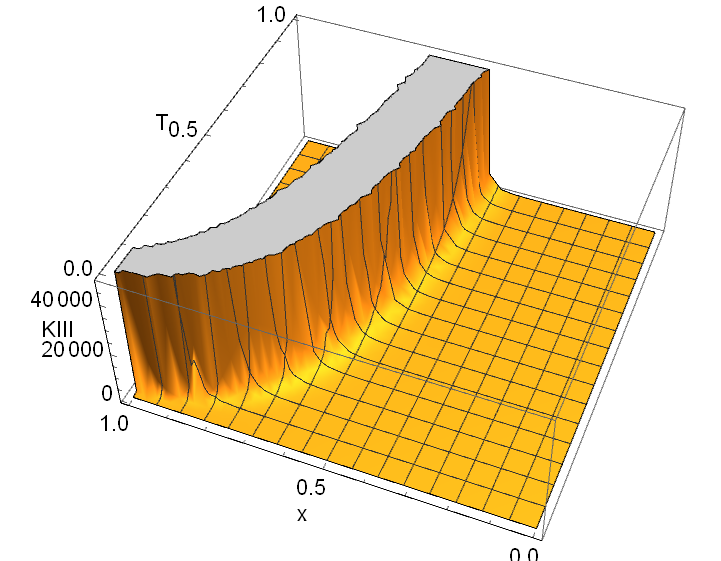}
\caption{Surface diagram $K_{III}$.}
\label{QG40}
\end{figure}
\begin{figure}[htbp]
\centering
		\includegraphics[width=0.7\textwidth]{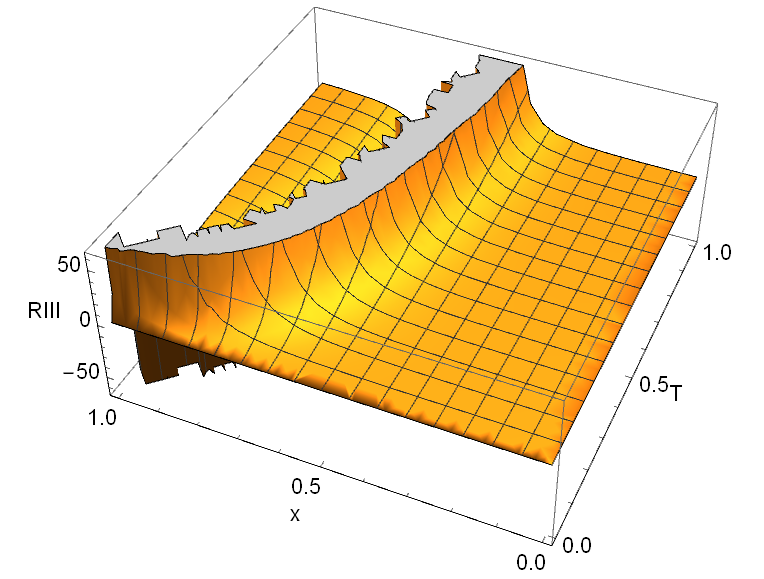}
\caption{Surface diagram $R_{III}$.}
\label{QG50}
\end{figure}
\clearpage
\section*{Bibliography} %
\bibliographystyle{apsrev4-2}
\bibliography{bibtestA}

\begin{thebibliography}{27}%
\makeatletter
\providecommand \@ifxundefined [1]{%
 \@ifx{#1\undefined}
}%
\providecommand \@ifnum [1]{%
 \ifnum #1\expandafter \@firstoftwo
 \else \expandafter \@secondoftwo
 \fi
}%
\providecommand \@ifx [1]{%
 \ifx #1\expandafter \@firstoftwo
 \else \expandafter \@secondoftwo
 \fi
}%
\providecommand \natexlab [1]{#1}%
\providecommand \enquote  [1]{``#1''}%
\providecommand \bibnamefont  [1]{#1}%
\providecommand \bibfnamefont [1]{#1}%
\providecommand \citenamefont [1]{#1}%
\providecommand \href@noop [0]{\@secondoftwo}%
\providecommand \href [0]{\begingroup \@sanitize@url \@href}%
\providecommand \@href[1]{\@@startlink{#1}\@@href}%
\providecommand \@@href[1]{\endgroup#1\@@endlink}%
\providecommand \@sanitize@url [0]{\catcode `\\12\catcode `\$12\catcode
  `\&12\catcode `\#12\catcode `\^12\catcode `\_12\catcode `\%12\relax}%
\providecommand \@@startlink[1]{}%
\providecommand \@@endlink[0]{}%
\providecommand \url  [0]{\begingroup\@sanitize@url \@url }%
\providecommand \@url [1]{\endgroup\@href {#1}{\urlprefix }}%
\providecommand \urlprefix  [0]{URL }%
\providecommand \Eprint [0]{\href }%
\providecommand \doibase [0]{https://doi.org/}%
\providecommand \selectlanguage [0]{\@gobble}%
\providecommand \bibinfo  [0]{\@secondoftwo}%
\providecommand \bibfield  [0]{\@secondoftwo}%
\providecommand \translation [1]{[#1]}%
\providecommand \BibitemOpen [0]{}%
\providecommand \bibitemStop [0]{}%
\providecommand \bibitemNoStop [0]{.\EOS\space}%
\providecommand \EOS [0]{\spacefactor3000\relax}%
\providecommand \BibitemShut  [1]{\csname bibitem#1\endcsname}%
\let\auto@bib@innerbib\@empty
\bibitem [{\citenamefont {Stanley}\ \emph {et~al.}(1999)\citenamefont
  {Stanley}, \citenamefont {Amaral}, \citenamefont {Canning}, \citenamefont
  {Gopikrishnan}, \citenamefont {Lee},\ and\ \citenamefont
  {Liu}}]{stanley1999}%
  \BibitemOpen
  \bibfield  {author} {\bibinfo {author} {\bibfnamefont {H.~E.}\ \bibnamefont
  {Stanley}}, \bibinfo {author} {\bibfnamefont {L.~N.}\ \bibnamefont {Amaral}},
  \bibinfo {author} {\bibfnamefont {D.}~\bibnamefont {Canning}}, \bibinfo
  {author} {\bibfnamefont {P.}~\bibnamefont {Gopikrishnan}}, \bibinfo {author}
  {\bibfnamefont {Y.}~\bibnamefont {Lee}},\ and\ \bibinfo {author}
  {\bibfnamefont {Y.}~\bibnamefont {Liu}},\ }\href@noop {} {\bibfield
  {journal} {\bibinfo  {journal} {Physica A: Statistical Mechanics and its
  Applications}\ }\textbf {\bibinfo {volume} {269}},\ \bibinfo {pages} {156}
  (\bibinfo {year} {1999})}\BibitemShut {NoStop}%
\bibitem [{\citenamefont {Čiegis}\ and\ \citenamefont
  {Čiegis}(2008)}]{Ciegis2008}%
  \BibitemOpen
  \bibfield  {author} {\bibinfo {author} {\bibfnamefont {R.}~\bibnamefont
  {Čiegis}}\ and\ \bibinfo {author} {\bibfnamefont {R.}~\bibnamefont
  {Čiegis}},\ }\href@noop {} {\bibfield  {journal} {\bibinfo  {journal}
  {Engineering Economics}\ }\textbf {\bibinfo {volume} {57}},\ \bibinfo {pages}
  {15} (\bibinfo {year} {2008})}\BibitemShut {NoStop}%
\bibitem [{\citenamefont {Niño}(2016)}]{Santos2016}%
  \BibitemOpen
  \bibfield  {author} {\bibinfo {author} {\bibfnamefont {A.~S.}\ \bibnamefont
  {Niño}},\ }\emph {\bibinfo {title} {Estudio del Papel del Ahorro en las
  Distribuciones de Dinero y Riqueza Mediante Herramientas de la Física
  Teórica}},\ \href@noop {} {\bibinfo {type} {Tesis de maestría}},\ \bibinfo
  {school} {Universidad Nacional de Colombia}, \bibinfo {address} {Bogotá,
  Colombia} (\bibinfo {year} {2016})\BibitemShut {NoStop}%
\bibitem [{\citenamefont {Rawlings}\ \emph
  {et~al.}(2004{\natexlab{a}})\citenamefont {Rawlings}, \citenamefont
  {Reguera},\ and\ \citenamefont {Reiss}}]{rawlings2004}%
  \BibitemOpen
  \bibfield  {author} {\bibinfo {author} {\bibfnamefont {P.~K.}\ \bibnamefont
  {Rawlings}}, \bibinfo {author} {\bibfnamefont {D.}~\bibnamefont {Reguera}},\
  and\ \bibinfo {author} {\bibfnamefont {H.}~\bibnamefont {Reiss}},\
  }\href@noop {} {\bibfield  {journal} {\bibinfo  {journal} {Physica A:
  Statistical Mechanics and its Applications}\ }\textbf {\bibinfo {volume}
  {343}},\ \bibinfo {pages} {643} (\bibinfo {year}
  {2004}{\natexlab{a}})}\BibitemShut {NoStop}%
\bibitem [{\citenamefont {Quevedo}\ and\ \citenamefont
  {Quevedo}(2011)}]{Quevedo2011}%
  \BibitemOpen
  \bibfield  {author} {\bibinfo {author} {\bibfnamefont {H.}~\bibnamefont
  {Quevedo}}\ and\ \bibinfo {author} {\bibfnamefont {M.~N.}\ \bibnamefont
  {Quevedo}},\ }\href {https://doi.org/https://doi.org/10.1155/2011/676495}
  {\bibfield  {journal} {\bibinfo  {journal} {Journal of Thermodynamics}\
  }\textbf {\bibinfo {volume} {2011}},\ \bibinfo {pages} {676495} (\bibinfo
  {year} {2011})},\ \Eprint
  {https://arxiv.org/abs/https://onlinelibrary.wiley.com/doi/pdf/10.1155/2011/676495}
  {https://onlinelibrary.wiley.com/doi/pdf/10.1155/2011/676495} \BibitemShut
  {NoStop}%
\bibitem [{\citenamefont {Quevedo}\ and\ \citenamefont
  {Quevedo}(2023)}]{Quevedo2023}%
  \BibitemOpen
  \bibfield  {author} {\bibinfo {author} {\bibfnamefont {H.}~\bibnamefont
  {Quevedo}}\ and\ \bibinfo {author} {\bibfnamefont {M.~N.}\ \bibnamefont
  {Quevedo}},\ }\href {https://doi.org/10.1142/S0219887823500573} {\bibfield
  {journal} {\bibinfo  {journal} {International Journal of Geometric Methods in
  Modern Physics}\ }\textbf {\bibinfo {volume} {20}},\ \bibinfo {pages}
  {2350057} (\bibinfo {year} {2023})},\ \Eprint
  {https://arxiv.org/abs/https://doi.org/10.1142/S0219887823500573}
  {https://doi.org/10.1142/S0219887823500573} \BibitemShut {NoStop}%
\bibitem [{\citenamefont {{Departamento Administrativo Nacional de Estadística
  (DANE)}}(2023)}]{dane2023}%
  \BibitemOpen
  \bibfield  {author} {\bibinfo {author} {\bibnamefont {{Departamento
  Administrativo Nacional de Estadística (DANE)}}},\ }\href
  {https://www.dane.gov.co/index.php/estadisticas-por-tema/cuentas-nacionales/cuentas-satelite/cuenta-satelite-del-deporte-de-bogota-csdb}
  {\bibinfo {title} {Cuenta satélite del deporte de bogotá (csdb)}} (\bibinfo
  {year} {2023}),\ \bibinfo {note} {información actualizada el 22 de junio de
  2023}\BibitemShut {NoStop}%
\bibitem [{\citenamefont {Quevedo~Cubillos}\ and\ \citenamefont
  {Quevedo}(2016)}]{Quevedo2016}%
  \BibitemOpen
  \bibfield  {author} {\bibinfo {author} {\bibfnamefont {H.}~\bibnamefont
  {Quevedo~Cubillos}}\ and\ \bibinfo {author} {\bibfnamefont {M.~N.}\
  \bibnamefont {Quevedo}},\ }\href
  {https://doi.org/10.15446/cuad.econ.v35n69.44876} {\bibfield  {journal}
  {\bibinfo  {journal} {Cuadernos de Economía}\ }\textbf {\bibinfo {volume}
  {35}},\ \bibinfo {pages} {691–707} (\bibinfo {year} {2016})}\BibitemShut
  {NoStop}%
\bibitem [{\citenamefont {Arya}\ and\ \citenamefont
  {Lardner}(1997)}]{arya1997}%
  \BibitemOpen
  \bibfield  {author} {\bibinfo {author} {\bibfnamefont {J.~C.}\ \bibnamefont
  {Arya}}\ and\ \bibinfo {author} {\bibfnamefont {R.~W.}\ \bibnamefont
  {Lardner}},\ }\href@noop {} {\emph {\bibinfo {title} {Matemáticas
  Aplicadas}}},\ \bibinfo {edition} {5th}\ ed.\ (\bibinfo  {publisher}
  {Prentice Hall},\ \bibinfo {address} {Ciudad de México},\ \bibinfo {year}
  {1997})\BibitemShut {NoStop}%
\bibitem [{\citenamefont {{Alcaldía Mayor de Bogotá}}(2023)}]{bogota2023}%
  \BibitemOpen
  \bibfield  {author} {\bibinfo {author} {\bibnamefont {{Alcaldía Mayor de
  Bogotá}}},\ }\href {https://bogota.gov.co/} {\bibinfo {title} {Portal web
  oficial de bogotá}} (\bibinfo {year} {2023}),\ \bibinfo {note} {accedido el
  1 de agosto de 2024}\BibitemShut {NoStop}%
\bibitem [{\citenamefont {{Instituto Distrital de Recreación y Deporte
  (IDRD)}}(2023)}]{idrd2023}%
  \BibitemOpen
  \bibfield  {author} {\bibinfo {author} {\bibnamefont {{Instituto Distrital de
  Recreación y Deporte (IDRD)}}},\ }\href {https://www.idrd.gov.co/} {\bibinfo
  {title} {Portal web oficial del instituto distrital de recreación y deporte
  (idrd)}} (\bibinfo {year} {2023}),\ \bibinfo {note} {accedido el 1 de agosto
  de 2024}\BibitemShut {NoStop}%
\bibitem [{\citenamefont {Dowling}(1997)}]{dowling1997}%
  \BibitemOpen
  \bibfield  {author} {\bibinfo {author} {\bibfnamefont {E.~T.}\ \bibnamefont
  {Dowling}},\ }\href@noop {} {\emph {\bibinfo {title} {Cálculo para
  administración, economía y ciencias sociales}}},\ \bibinfo {edition} {4th}\
  ed.\ (\bibinfo  {publisher} {McGraw-Hill},\ \bibinfo {address} {Ciudad de
  México},\ \bibinfo {year} {1997})\BibitemShut {NoStop}%
\bibitem [{\citenamefont {Hoy}\ \emph {et~al.}(2001)\citenamefont {Hoy},
  \citenamefont {Livernois}, \citenamefont {McKenna}, \citenamefont {Rees},\
  and\ \citenamefont {Stengos}}]{hoy2001}%
  \BibitemOpen
  \bibfield  {author} {\bibinfo {author} {\bibfnamefont {M.}~\bibnamefont
  {Hoy}}, \bibinfo {author} {\bibfnamefont {J.}~\bibnamefont {Livernois}},
  \bibinfo {author} {\bibfnamefont {C.}~\bibnamefont {McKenna}}, \bibinfo
  {author} {\bibfnamefont {R.}~\bibnamefont {Rees}},\ and\ \bibinfo {author}
  {\bibfnamefont {T.}~\bibnamefont {Stengos}},\ }\href@noop {} {\emph {\bibinfo
  {title} {Mathematics for Economics}}},\ \bibinfo {edition} {2nd}\ ed.\
  (\bibinfo  {publisher} {MIT Press},\ \bibinfo {address} {Cambridge, MA},\
  \bibinfo {year} {2001})\BibitemShut {NoStop}%
\bibitem [{\citenamefont {Hoffmann}\ and\ \citenamefont
  {Bradley}(2007)}]{hoffmann2007}%
  \BibitemOpen
  \bibfield  {author} {\bibinfo {author} {\bibfnamefont {L.~D.}\ \bibnamefont
  {Hoffmann}}\ and\ \bibinfo {author} {\bibfnamefont {G.~L.}\ \bibnamefont
  {Bradley}},\ }\href@noop {} {\emph {\bibinfo {title} {Calculus for Business,
  Economics, and the Social and Life Sciences}}},\ \bibinfo {edition} {10th}\
  ed.\ (\bibinfo  {publisher} {McGraw-Hill},\ \bibinfo {address} {New York},\
  \bibinfo {year} {2007})\BibitemShut {NoStop}%
\bibitem [{\citenamefont {Santos Ni\~{n}o}(2016)}]{Santos_2016}%
  \BibitemOpen
  \bibfield  {author} {\bibinfo {author} {\bibfnamefont {A.}~\bibnamefont
  {Santos Ni\~{n}o}},\ }\href@noop {} {\emph {\bibinfo {title} {Tesis de
  Maestr{\'a}: Estudio del papel del ahorro en las distribuciones de dinero y
  riqueza mediante herramientas de la f{\'i}sica te{\'o}rica}}}\ (\bibinfo
  {publisher} {Departamento de F{\'isica}. Universidad Nacional de Colombia.
  Director: Quimbay Herrera, Carlos Jos{\'e}},\ \bibinfo {year}
  {2016})\BibitemShut {NoStop}%
\bibitem [{\citenamefont {Rawlings}\ \emph
  {et~al.}(2004{\natexlab{b}})\citenamefont {Rawlings}, \citenamefont
  {Reguera},\ and\ \citenamefont {Reiss}}]{RAWLINGS2004643}%
  \BibitemOpen
  \bibfield  {author} {\bibinfo {author} {\bibfnamefont {P.~K.}\ \bibnamefont
  {Rawlings}}, \bibinfo {author} {\bibfnamefont {D.}~\bibnamefont {Reguera}},\
  and\ \bibinfo {author} {\bibfnamefont {H.}~\bibnamefont {Reiss}},\ }\href
  {https://doi.org/https://doi.org/10.1016/j.physa.2004.06.152} {\bibfield
  {journal} {\bibinfo  {journal} {Physica A: Statistical Mechanics and its
  Applications}\ }\textbf {\bibinfo {volume} {343}},\ \bibinfo {pages} {643 }
  (\bibinfo {year} {2004}{\natexlab{b}})}\BibitemShut {NoStop}%
\bibitem [{\citenamefont {Alonso}\ \emph {et~al.}(1976)\citenamefont {Alonso},
  \citenamefont {Finn}, \citenamefont {Heras},\ and\ \citenamefont
  {Araujo}}]{alonso1976fisica}%
  \BibitemOpen
  \bibfield  {author} {\bibinfo {author} {\bibfnamefont {M.}~\bibnamefont
  {Alonso}}, \bibinfo {author} {\bibfnamefont {E.}~\bibnamefont {Finn}},
  \bibinfo {author} {\bibfnamefont {C.}~\bibnamefont {Heras}},\ and\ \bibinfo
  {author} {\bibfnamefont {J.}~\bibnamefont {Araujo}},\ }\href
  {https://books.google.com.co/books?id=wcWxAAAACAAJ} {\emph {\bibinfo {title}
  {F{\'\i}sica, vol. III: fundamentos cu{\'a}nticos y estad{\'\i}sticos}}},\
  F{\'\i}sica\ (\bibinfo  {publisher} {Pearson-Educaci{\'o}n},\ \bibinfo {year}
  {1976})\BibitemShut {NoStop}%
\bibitem [{\citenamefont {Reif}(1996)}]{reif1996fisica}%
  \BibitemOpen
  \bibfield  {author} {\bibinfo {author} {\bibfnamefont {F.}~\bibnamefont
  {Reif}},\ }\href {https://books.google.hn/books?id=ygGc7I\_9MWEC} {\emph
  {\bibinfo {title} {F{\'\i}sica estad{\'\i}stica}}},\ \bibinfo {series}
  {Berkeley physics course}\ No.\ \bibinfo {number} {v. 5}\ (\bibinfo
  {publisher} {Revert{\'e}},\ \bibinfo {year} {1996})\BibitemShut {NoStop}%
\bibitem [{\citenamefont {Landau}\ and\ \citenamefont
  {Lifshitz}(1988)}]{landau1988fisica}%
  \BibitemOpen
  \bibfield  {author} {\bibinfo {author} {\bibfnamefont {L.}~\bibnamefont
  {Landau}}\ and\ \bibinfo {author} {\bibfnamefont {E.}~\bibnamefont
  {Lifshitz}},\ }\href {https://books.google.com.co/books?id=ViiLj9kchoUC}
  {\emph {\bibinfo {title} {F{\'\i}sica estad{\'\i}stica}}},\ \bibinfo {series}
  {Ciencias Qu{\'\i}micas: Ingenier{\'\i}a qu{\'\i}mica}\ No.\ \bibinfo
  {number} {v. 5}\ (\bibinfo  {publisher} {Revert{\'e}},\ \bibinfo {year}
  {1988})\BibitemShut {NoStop}%
\bibitem [{\citenamefont {Greiner}\ \emph {et~al.}(2012)\citenamefont
  {Greiner}, \citenamefont {Rischke}, \citenamefont {Neise},\ and\
  \citenamefont {St{\"o}cker}}]{greiner2012thermodynamics}%
  \BibitemOpen
  \bibfield  {author} {\bibinfo {author} {\bibfnamefont {W.}~\bibnamefont
  {Greiner}}, \bibinfo {author} {\bibfnamefont {D.}~\bibnamefont {Rischke}},
  \bibinfo {author} {\bibfnamefont {L.}~\bibnamefont {Neise}},\ and\ \bibinfo
  {author} {\bibfnamefont {H.}~\bibnamefont {St{\"o}cker}},\ }\href
  {https://books.google.com.co/books?id=8jrTBwAAQBAJ} {\emph {\bibinfo {title}
  {Thermodynamics and Statistical Mechanics}}},\ Classical Theoretical Physics\
  (\bibinfo  {publisher} {Springer New York},\ \bibinfo {year}
  {2012})\BibitemShut {NoStop}%
\bibitem [{\citenamefont {Quevedo}(2008)}]{Quevedo2007}%
  \BibitemOpen
  \bibfield  {author} {\bibinfo {author} {\bibfnamefont {H.}~\bibnamefont
  {Quevedo}},\ }\href {https://doi.org/10.1007/s10714-007-0586-0} {\bibfield
  {journal} {\bibinfo  {journal} {Gen. Rel. Grav.}\ }\textbf {\bibinfo {volume}
  {40}},\ \bibinfo {pages} {971} (\bibinfo {year} {2008})},\ \Eprint
  {https://arxiv.org/abs/0704.3102} {arXiv:0704.3102 [gr-qc]} \BibitemShut
  {NoStop}%
\bibitem [{\citenamefont {Larranaga}\ and\ \citenamefont
  {Cardenas}(2012)}]{Larranaga2011}%
  \BibitemOpen
  \bibfield  {author} {\bibinfo {author} {\bibfnamefont {A.}~\bibnamefont
  {Larranaga}}\ and\ \bibinfo {author} {\bibfnamefont {A.}~\bibnamefont
  {Cardenas}},\ }\href {https://doi.org/10.3938/jkps.60.987} {\bibfield
  {journal} {\bibinfo  {journal} {J. Korean Phys. Soc.}\ }\textbf {\bibinfo
  {volume} {60}},\ \bibinfo {pages} {987} (\bibinfo {year} {2012})},\ \Eprint
  {https://arxiv.org/abs/1108.2205} {arXiv:1108.2205 [gr-qc]} \BibitemShut
  {NoStop}%
\bibitem [{\citenamefont {Porras}(2016)}]{valdes2016interpretacion}%
  \BibitemOpen
  \bibfield  {author} {\bibinfo {author} {\bibfnamefont {E.~A.~V.}\
  \bibnamefont {Porras}},\ }\emph {\bibinfo {title} {Interpretación
  estadística de las métricas geometrotermodinámicas}},\ \href@noop {}
  {\bibinfo {type} {Tesis de maestría en ciencias físicas}},\ \bibinfo
  {school} {Universidad Nacional Autónoma de México}, \bibinfo {address}
  {Ciudad de México, México} (\bibinfo {year} {2016}),\ \bibinfo {note}
  {instituto de Ciencias Nucleares, Tutor: Dr. Hernando Quevedo
  Cubillos}\BibitemShut {NoStop}%
\bibitem [{\citenamefont {Reyes}(2019)}]{pineda2019geometrotermodinamica}%
  \BibitemOpen
  \bibfield  {author} {\bibinfo {author} {\bibfnamefont {V.~P.}\ \bibnamefont
  {Reyes}},\ }\emph {\bibinfo {title} {Geometrotermodinámica Estadística}},\
  \href@noop {} {\bibinfo {type} {Tesis de doctorado en ciencias (física)}},\
  \bibinfo  {school} {Universidad Nacional Autónoma de México}, \bibinfo
  {address} {Ciudad de México, México} (\bibinfo {year} {2019}),\ \bibinfo
  {note} {instituto de Ciencias Nucleares, Director: Dr. Hernando Quevedo
  Cubillos}\BibitemShut {NoStop}%
\bibitem [{\citenamefont {Spiegel}\ and\ \citenamefont
  {Liu}(1999)}]{spiegel1999mathematical}%
  \BibitemOpen
  \bibfield  {author} {\bibinfo {author} {\bibfnamefont {M.}~\bibnamefont
  {Spiegel}}\ and\ \bibinfo {author} {\bibfnamefont {J.}~\bibnamefont {Liu}},\
  }\href {https://books.google.com.co/books?id=jIMHMX6iUYIC} {\emph {\bibinfo
  {title} {Mathematical Handbook of Formulas and Tables}}},\ Schaum's outline
  of theory and problems\ (\bibinfo  {publisher} {McGraw-Hill},\ \bibinfo
  {year} {1999})\BibitemShut {NoStop}%
\bibitem [{\citenamefont {contributors}(2024)}]{wikipedia2024pandemia}%
  \BibitemOpen
  \bibfield  {author} {\bibinfo {author} {\bibfnamefont {W.}~\bibnamefont
  {contributors}},\ }\href {https://es.wikipedia.org/wiki/Pandemia_de_COVID-19}
  {\bibinfo {title} {Pandemia de covid-19 --- wikipedia, la enciclopedia
  libre}} (\bibinfo {year} {2024}),\ \bibinfo {note} {[En línea; consultado el
  7-octubre-2024]}\BibitemShut {NoStop}%
\bibitem [{\citenamefont {Quevedo}\ and\ \citenamefont
  {Sanchez}(2008)}]{Quevedo2008xn}%
  \BibitemOpen
  \bibfield  {author} {\bibinfo {author} {\bibfnamefont {H.}~\bibnamefont
  {Quevedo}}\ and\ \bibinfo {author} {\bibfnamefont {A.}~\bibnamefont
  {Sanchez}},\ }\href {https://doi.org/10.1088/1126-6708/2008/09/034}
  {\bibfield  {journal} {\bibinfo  {journal} {JHEP}\ }\textbf {\bibinfo
  {volume} {09}},\ \bibinfo {pages} {034}}\BibitemShut {NoStop}%
\end{thebibliography}%

\end{document}